\documentclass[]{fairmeta}

\title{PersonaMem-v3: Toward Omni-Platform Personal Intelligence for Holistic User Understanding, Recommendation, and Agentic Tasks}

\author[2]{Bowen Jiang}
\author[2]{Yuan Yuan}
\author[2]{Zhuoqun Hao}
\author[1]{Yuchen Liu}
\author[3]{Maohao Shen}
\author[2]{Sihao Chen}
\author[3]{Gregory Wornell}
\author[2]{Chris Callison-Burch}
\author[2]{Lyle Ungar}
\author[2]{Dan Roth}
\author[1]{Iordanis Fostiropoulos}
\author[1]{Qi Guo}
\author[1]{Xiangjun Fan}
\author[2]{\\Camillo J. Taylor}
\author[1,\dagger]{Hanchao Yu}

\affiliation[1]{Meta Recommendation Systems}
\affiliation[2]{University of Pennsylvania}
\affiliation[3]{MIT}

\abstract{Personal intelligence is becoming a central frontier for user-facing AI agents. To be helpful in everyday life, agents must understand users across the digital contexts where their preferences, intents, habits, social relationships, and needs unfold over time. Today’s systems can personalize within individual apps or tasks, but personal intelligence as a whole remains under-measured: how agents build cross-context user understanding, support steerable recommendation systems, act proactively across platforms, and avoid over-personalization. We introduce PersonaMem-v3, a real-world-grounded benchmark and evaluation harness for omni-platform personal intelligence. PersonaMem-v3 is seeded from more than one million anonymized real-world engagement histories, most of which are implicit signals, and uses them to construct time-indexed user digital worlds across social media, chatbot, calendar, and AI-companion with preference evolvement over time. The benchmark brings personalization, LLM-powered recommendation, proactiveness, agentic tool use, and geo-temporal reasoning into one framework, anchored in psychology, social-linguistics, and user-behavior theories. It evaluates whether AI agents can infer holistic user understanding from cross-platform evidence, personalize responses, rerank recommendations on social media, follow user steering through natural language, and hold back when personalization would be inappropriate, repetitive, outdated, or unnecessary. PersonaMem-v3 points toward LLM-powered personal intelligent agents that work with existing scalable recommendation infrastructure while making personalization more interactive, agentic, and aligned with how real users experience their digital lives.
}

\contribution[\dagger]{Project Lead}

\date{August 25, 2026}
\correspondence{Bowen Jiang <\email{bwjiang@seas.upenn.edu}> and Hanchao Yu <\email{hanchaoyu@meta.com}>}

\metadata[Data]{\url{https://huggingface.co/datasets/bowen-upenn/PersonaMem-v3}}
\metadata[Code]{\url{https://github.com/bowen-upenn/PersonaMem-v3}}
\metadata[Prior work]{\url{https://arxiv.org/abs/2504.14225} (v1); \url{https://arxiv.org/abs/2512.06688} (v2)}

\usepackage{longtable}

\usepackage{afterpage}

\begin{document}

\maketitle

\afterpage{%
  \begin{figure}[!t]
    \centering
    \includegraphics[width=\linewidth]{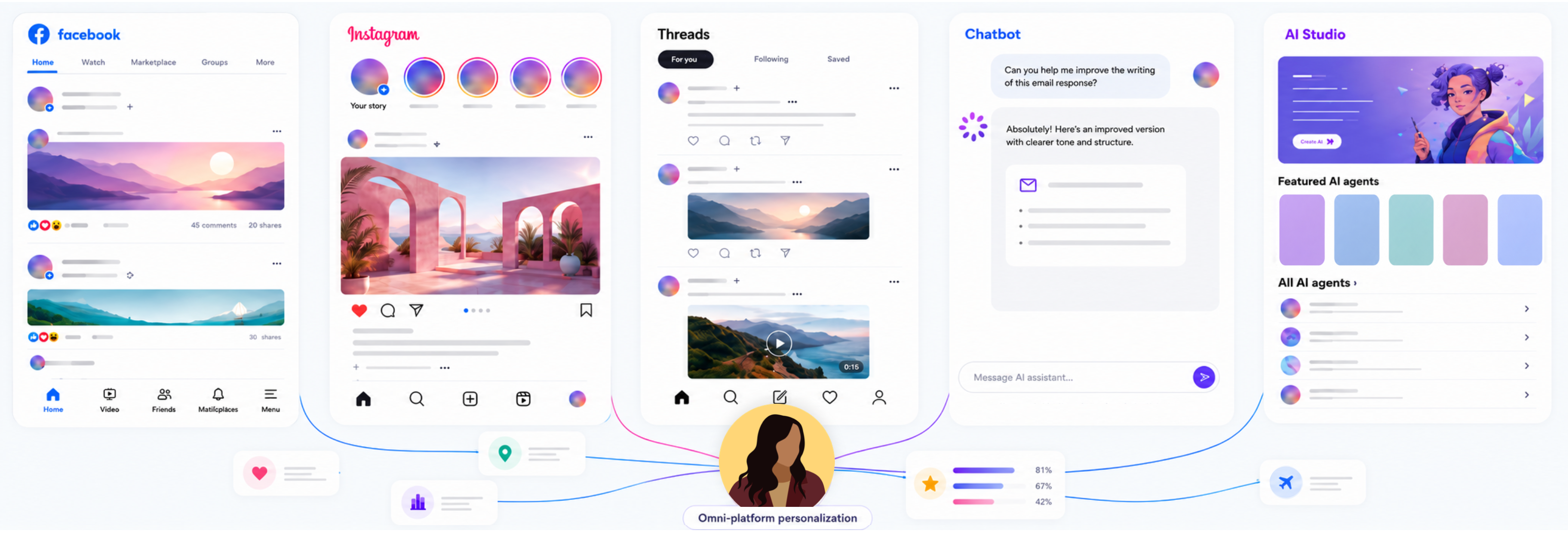}
    \caption{PersonaMem-v3 studies omni-platform personalization: how a system can understand users across the many platforms where they interact online, including social media and chatbots. It uses users’ engagement histories to build holistic understanding of their personas and preferences, enabling personalized and socially appropriate responses and agentic task completion. We seed PersonaMem-v3 from real-world, large-scale, anonymized, and privacy-preserving data, and synthesize rich multi-platform user histories in textual-only annotations to mimic this digital environment. This figure is illustrative only and does not represent how each platform actually looks like.}
    \label{fig:header}
  \end{figure}%
}

\section{Introduction}
\label{section:intro}

Personal intelligence is becoming part of everyday life~\citep{ MetaSuperintelligence2025, meta2026musespark, google2026geminipersonalintelligence, openai2025personalizedai, openclaw2026, nousresearch2026hermesagent}. A personalized agent has to understand the whole person: what they care about, how their interests evolve over time, what they want recommended next, and when to stay out of the way in different time and scenarios. Across the digital contexts they use, people reveal different facets of their preferences and needs that no single system sees in full. AI chatbots~\citep{jiang2025know, jiang2025personamem, salemi2026pathways, maharana2024locomo, wu2024longmemeval, zhao2025prefeval} and social media's recommendation systems~\citep{niu2018neural, zhai2024actions, zhang2024scaling, zhang2024wukong, naumov2019deep, pei2019personalized, bello2018seq2slate, zhang2024linear, liu2025score, li2022pear, gao2025llm4rerank} each personalize within their own domain, but they operate independently, leaving their understanding of the user fragmented. The next step is personal intelligence that spans a user’s all-day digital life, builds on shared memory across these contexts, and acts on that fuller understanding in ways users find both helpful and comfortable.

Real users are complicated. Their personas are layered and platform-dependent, shaped by hidden motivations; their preferences shift over time; and most of what they reveal emerges implicitly through many small engagements, which purely synthetic-persona benchmarks cannot reproduce. PersonaMem-v3 is grounded two hundred anonymized real users and more than one million privacy-preserving social-media engagement events that cannot be traced back to actual individuals~\citep{fostiropoulos2026gistbench}. This shift is foundational: PersonaMem-v3 anchors every persona claim and predicted user preference in sufficient real engagement evidence from that individual. These logs span explicit signals such as likes, dislikes, saves, follows, comments, posts, reposts, shares, direct messages, as well as implicit signals such as lingering views, high watch-through, scroll-pasts, and non-responses in message threads. Only about five percent of the signal is explicit; the remaining ninety-five percent is implicit~\citep{fostiropoulos2026gistbench}.

PersonaMem-v3 evaluates \textbf{omni-platform personal intelligence} for real-world users beyond a single conversational channel. A user commonly moves between feeds, messaging, chatbots, and companion characters within the same day, and the picture of who they are only becomes coherent when these surfaces can be interpreted in relation to one another. PersonaMem-v3 operationalizes this cross-context view through six connected digital surfaces: Instagram, Facebook, Threads, Chatbot, AI Studio, and Calendar, each instantiated as a backend database. These surfaces are generated from the same underlying user persona and behavioral history, encouraging models to reason over shared context, such as using preferences learned from Chatbot interactions to personalize feed recommendations on Instagram. 

This setup also moves PersonaMem-v3 beyond static recommendation toward an interactable paradigm. Today’s recommender systems are largely one-way: the system feeds, the user consumes. \textbf{The next paradigm brings together chatbot personalization, neural recommendation, and personalized agentic tasks into one holistic surface}, where the assistant can both decide what to surface and act on the user’s behalf. The reranker behind classical recommendation becomes LLM-driven~\citep{gao2025llm4rerank, hou2024zeroshotrankers, luo2025recranker, liu2026generating, wang2025userfeedbackalignment, wang2024interestexploration}, and users can steer it directly through natural language by mentioning the in-feed AI to ask for more like this, stop a topic, or step outside the usual bubble. PersonaMem-v3 models this user-interactable, LLM-driven recommendation paradigm and treats personal intelligence as a whole-person-understanding problem rather than a feed-tuning one.

Personalization is fundamentally a nuanced model-behavior problem. Useful personalization requires more than recalling what a user once liked: it must account for behavioral challenges such as over-personalization~\citep{hu2026op, spadea2025avoiding}, short-term shifts in preferences and circumstances~\citep{jiang2025know, jiang2025personamem}, and the need to be proactive without becoming intrusive. In particular, we introduce the notion of the \textit{over-personalization tax}: as models become better at leveraging user context, especially when equipped with additional user memory modules, they might also become more likely to over-personalize. We capture four common forms of over-personalization: fatigue, inappropriateness, irrelevance, and sycophancy. A system may over-reinforce a user’s interests by invoking them too often, say something like ``I know you like X'' too explicitly, lecture the user, reveal sensitive or socially inappropriate inferences, or inject personal context into requests that do not actually need it. In addition, useful personalization must distinguish enduring preferences from transient ones, and track geographic and situational changes over time. Separately, it covers proactive personalization~\citep{chen2026vitabench2, lu2025proactiveagent, yang2025contextagent, sun2025training, tang2026proagentbench, zhang2026pibench}: knowing when to step in and when to hold back in agentic tasks, from surfacing unread close-friend messages to issuing timely alerts when evolving context suggests the user may need help.

PersonaMem-v3 turns these challenges into an evaluation harness that no longer treats test queries independently, but places them within a temporally evolving history of user-AI activities, with support for tool calling and agentic search~\citep{anthropic2026claudecodetools} for AI agents to interact with backend databases. To make our personalization targets more realistic, PersonaMem-v3 builds personas, preferences, and voices from behaviorally grounded histories, and constrains its persona generation and automatic validation with over twenty named frameworks from psychology, sociology, linguistics, behavioral science, and human-computer interaction. This lets the evaluation harness test not only whether a model can retrieve user context, but whether it can use that context coherently, appropriately, and without over-personalizing.

\paragraph{Main contributions:}
\begin{itemize}
  \item We introduce PersonaMem-v3, a state-of-the-art personalization benchmark grounded in million-scale anonymized real-world user activities, where every persona claim traces to multiple real engagements.
  \item We construct omni-platform user worlds spanning six platforms, including social media and chatbots, including a pipeline that converts large-scale, noisy, raw engagement history into holistic user profiling.
  \item We present a psychology-grounded user modeling and evaluation harness that measures personalization, over-personalization, proactiveness, recommendation, and agentic tasks for LLMs and AI agents.
  \item Our experiments show that current agents still solve a limited portion of omni-platform personal intelligence, with recommendation reranking and proactive actions remaining especially challenging. We find that user-steerable recommendation through in-feed natural-language directives has a strong potential, memory modules substantially improve efficiency, but have model-dependent and even opposite effects on over-personalization, and self-evolving agentic memory still suffer from lost-in-the-middle failures when iteratively compressing long user histories.
\end{itemize}

\section{Benchmark Overview}
\label{section:benchmark}

A personalization benchmark should look like the world it is measuring. This section walks through how PersonaMem-v3 builds that world in three layers. We begin with the real interaction data we ground in. We then layer on the persona, preferences, voice, and motivation machinery that makes every persona claim event-anchored, cross-referenced, and psychology-grounded. We end with the realistic interaction data the pipeline produces across six social media and chatbot apps we model as each user's example of digital life.

\subsection{Real-world data foundation}
\label{subsec:data_foundation}
PersonaMem-v3 builds on Meta’s GIST-Bench~\citep{fostiropoulos2026gistbench}, a privacy-preserving dataset derived from real-world social-media engagements. GIST-Bench groups approximately every ten real users with overlapping interest profiles into one cohort, aggregates their engagement histories, and resamples from each pooled cohort to create synthetic users that preserve realistic behavioral distributions without retaining any individual user’s traceable trajectory. It further removes direct identifiers and original media links. 

PersonaMem-v3 randomly samples 200 users with enough history and over 1,000,000 engagement events from GIST-Bench, averaging more than 4,000 events per user over a 30-day observation window. Each raw event records an interaction with hashtag-described content, together with a timestamp, user ID, and one of four signal types: \textit{explicit positive, explicit negative, implicit positive, or implicit negative}, nearly 63.8\% of which are implicit negative, corresponding to cases where the user simply skips the content. The user ID allows us to trace each unique user’s full engagement history across events.
Together, these events provide a real-world behavioral substrate for studying personalization from sparse, noisy, and uneven engagement evidence rather than from fully synthetic persona descriptions and history.

\subsection{Back-engineering preferences from raw events}
\label{subsec:pref_inference}

GIST-Bench gives us a real but deliberately thin substrate: hashtag-anchored engagement rows and timestamps. There is no self-description, no demographic field, no narrative, no voice.
PersonaMem-v3 extends this substrate into complete user worlds, synthesizing richer personas, preferences, voices, and timestamped interaction histories across multiple platforms with the OpenAI's GPT-5.5~\citep{openai2026gpt55systemcard} API. Because this expansion adds structure beyond the raw rows, reliability is the central constraint: the pipeline must enrich the substrate without inventing claims the evidence cannot support. We address this through a three-stage curation pipeline: candidate preferences are first extracted from individual events and retained only if they survive cross-validation; the surviving preferences are then assembled into coherent layered personas and audited; finally, the validated user model is distributed across six platforms as realistic timestamped events. Throughout, every synthesized persona, preference, and interaction will be grounded in multiple original events from the real world user data.

\subsubsection{From noisy engagement to grounded preferences}
\label{subsubsec:atomic_inference}

\begin{figure}[!t]
    \centering
    \includegraphics[width=\linewidth]{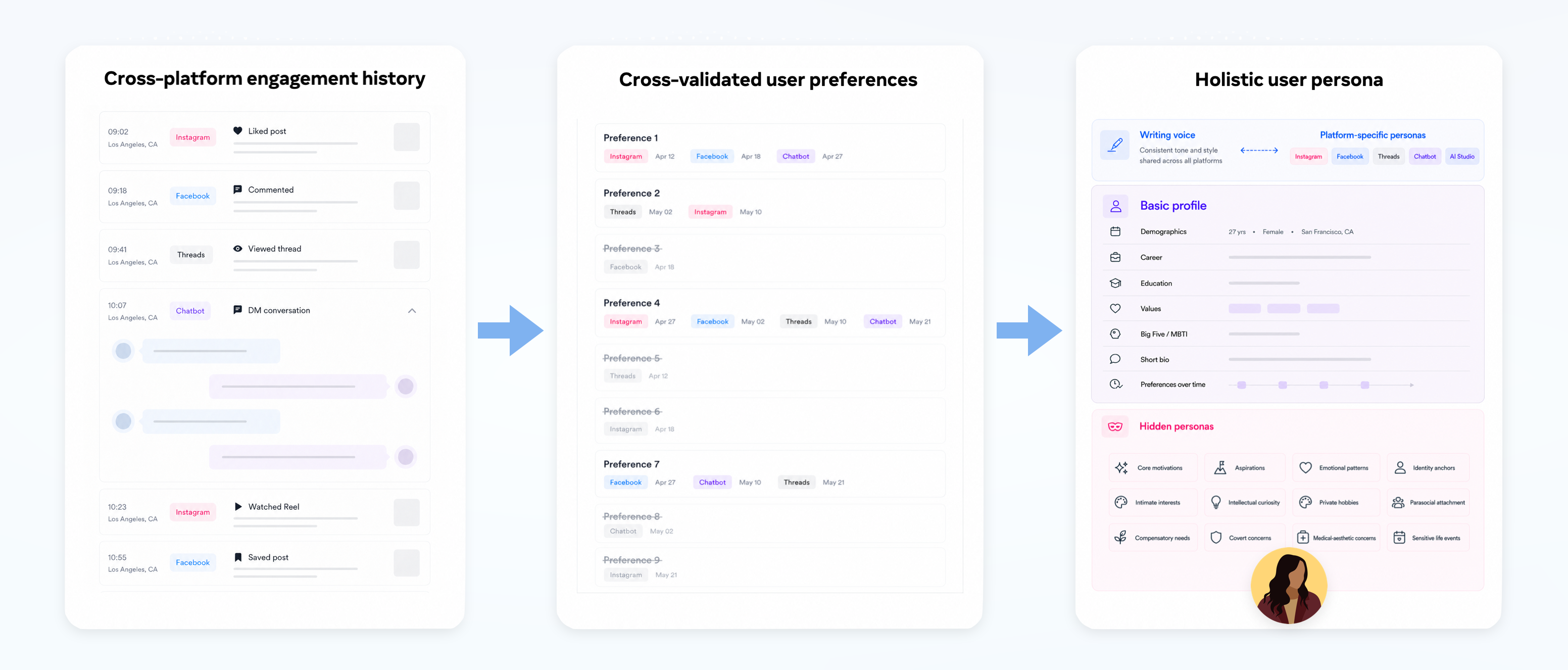}
    \caption{Data generation pipeline of PersonaMem-v3. Starting from cross-platform engagement histories from the real world, PersonaMem-v3 infers candidate preferences from noisy user actions, cross-validates them across repeated evidence, prunes weak claims, and organizes the surviving signals into a holistic user persona. The final persona includes a basic profile, a writing voice, platform-specific personas, and hidden personas as deep motivations, enabling benchmark users to have grounded, holistic, and realistic multi-platform digital lives and persona profiles.}
    \label{fig:data_flow}
  \end{figure}%

A single engagement rarely tells us exactly what a user likes. This is especially true for implicit negatives, where the user quickly scrolls past content. A skip may mean dislike, but it may also mean the user was busy, distracted, or simply scrolling quickly. PersonaMem-v3 therefore never infer persona claims directly from a single scroll-past. Instead, it first focuses on cleaner signals: explicit positives, explicit negatives, and implicit positives. For each such event, PersonaMem-v3 predicts up to three candidate atomic preferences from different plausible angles, grounds each candidate in the event’s hashtags, and assigns it an initial confidence score. These candidates are possible readings of behavior. For implicit negative ones, we only upgrades the signal when repetition makes the pattern meaningful. Many similar scroll-pasts in a short window suggest the user may actively not want that content, and we promote strong repeated patterns into explicit negative preferences. Otherwise, skipped events remain in the behavioral history but do not shape the persona.

The candidate preferences are then merged, categorized, and cross-validated across the user’s history. Since each event may be interpreted from multiple plausible angles, the initial candidates intentionally over-hallucinate possible readings of the user. The validation step then asks which of these readings are supported repeatedly across independent engagements. Repeated evidence strengthens a claim, contradictory evidence weakens it, and explicit actions count more than implicit ones because they are more reliable. We measure this heuristically with a cross-reference score: each explicit corroborating event contributes 1 point, while each implicit corroborating event contributes 0.5 points. To enter the final user profile, a preference must pass both an initial confidence floor and an evidence-support threshold. Positive claims require at least 0.75 initial confidence, while negative claims require at least 0.55. For long-term preferences, the cross-reference threshold is 20 when the evidence is explicit and 50 when it is implicit. Short-term intents use a lower threshold of 3 because they naturally have less time to accumulate repeated evidence. Claims with weak or inconsistent support are pruned. We find these dataset-specific values empirically. In this way, PersonaMem-v3 turns many noisy event-level interpretations into a smaller, more grounded set of user preferences.

\subsubsection{Preference evolution over long context}
\label{subsubsec:contradictions_horizons}

Real user preferences are not static. They can strengthen, fade, become more specific, or reverse as the user’s context changes. PersonaMem-v3 uses timestamped real-world engagement histories to capture such nuanced evolution, rather than treating preferences as fixed labels or manually synthesizing preference shifts. Importantly, these shifts are often interdependent: a change in one preference may reflect broader changes in the user’s goals, life events, social context, platform usage, or newly emerging interests. By grounding preference evolution in real-world data, PersonaMem-v3 captures fine-grained user preference changes and their dependencies over time.

Conflicting signals are interpreted in long context. Each user history spans approximately 400,000 tokens on average, creating a realistic setting where isolated contradictions must be distinguished from broader temporal patterns. Isolated contradictions are treated as noise, while repeated patterns can indicate a meaningful shift in what the user wants. PersonaMem-v3 also separates enduring preferences from short-term intents: an interest in a hobby or aesthetic may persist, while interests tied to a trip, event, purchase, or medical consultation may expire once the situation ends. This makes long-context in-context learning a central real-world challenge for personalization~\citep{dou2026cl, jiang2025know, bai2025longbench, li2024long}, since agents must infer not only what the user likes, but when that preference is still relevant. The final preferences are therefore grounded claims that survive cross-validation, contradiction and temporal interpretation.

\subsection{Building coherent, holistic, and psychology-grounded personas}
\label{subsec:persona_aggregation}

A list of preferences is not yet a person. PersonaMem-v3 turns those preferences into a coherent persona by organizing them into aligned layers: a basic profile, a writing voice, app-specific self-presentations, hidden motivations sit beneath the surface engagement, and sensitive life context. This makes the persona holistic rather than just a flat list of likes and dislikes. It is also psychology-grounded: each higher-level layer is tied to a named framework and implemented through concrete schema fields, motivation frames, or validation gates, so the theory constrains the persona generation pipeline.

\subsubsection{The basic profile}
\label{subsubsec:basic_profile}

The basic profile gives each user a stable biographical frame. PersonaMem-v3 samples demographics from diversified distributions of race, gender, and sexual orientation, then uses the validated preference signals to generate a culturally consistent name, plausible career, education background, Big Five traits, MBTI, and short bio with reasonable randomness. These fields serve as narrative anchors, helping later generations keep the user’s communication style and behaviors roughly consistent.

\subsubsection{The writing voice and platform-specific personas}
\label{subsubsec:shared_voice}

Each person has a personalized writing voice: the recurring patterns in their wording, sentence structure, hedging, emotional tone, level of directness, preferred constructions, and the kinds of speech genres they naturally use. PersonaMem-v3 models this voice as a shared layer across all generated text, so the same user does not sound like different people at different times. This shared voice captures the user’s deep and stable expressive habits, rather than simply copying repeated phrases, emoji patterns, or surface-level word choices.

{\footnotesize
\begin{longtable}{p{0.16\textwidth}p{0.56\textwidth}p{0.22\textwidth}}
  \caption{The four layers of the shared user voice. Layers~1--3 stay coherent across apps. Per-app differences mainly appear in Layer~4, with each app selecting and lightly reweighting parts of Layer~3. A separate voice-avoid block prevents forbidden tones and repeated catchphrases.}
  \label{table:voice_layers} \\
  \toprule
  \textbf{Layer} & \textbf{Voice signal} & \textbf{Theoretical anchor} \\
  \midrule
  \endfirsthead

  \toprule
  \textbf{Layer} & \textbf{Voice signal} & \textbf{Theoretical anchor} \\
  \midrule
  \endhead

  \midrule
  \endfoot

  \bottomrule
  \endlastfoot

  1.\ Identity spine & What the user repeatedly brings into text: core themes, motives, emotional baseline, and personality drivers. Implemented through agency, redemption and contamination motifs, life-stage concerns, signature concerns, LIWC anchors, and Big-Five drivers. & McAdams~\citep{mcadams1985power}; LIWC~\citep{pennebaker2015development}; Big~Five~\citep{mccrae1992introduction} \\[2pt]

  2.\ Idiolect & How the user writes at the sentence level, without mechanically copying phrases. Implemented through function-word profile, sentence shape, appraisal style, abstract slot patterns, capitalization, punctuation, formality, and emoji palette. & Martin and White APPRAISAL~\citep{martin2005language}; Construction Grammar~\citep{goldberg1995constructions} \\[2pt]

  3.\ Indexical repertoire & Which expressive modes the user can naturally switch into while still sounding like the same person. Implemented through stances, registers, backstage/frontstage range, and speech-genre fluency. & Bakhtin speech-genre theory~\citep{bakhtin1986speech}; Goffman~\citep{goffman1959presentation} \\[2pt]

  4.\ Surface modulation & What changes because of app, audience, or context, rather than because the user became a different person. Implemented through length, emoji intensity, self-censoring, disclosure depth, topical focus, and posting rhythm. & Bell audience design~\citep{bell1984language} \\
\end{longtable}
}

Meanwhile, voice is not completely static. How a person writes can shift slightly depending on why they are using an app, who they expect to see the post, and what kind of social context the platform creates.
PersonaMem-v3 captures this through four per-app sub-personas, one each for Instagram, Facebook, Threads, and the AI Chatbot. Following Bell's audience design~\citep{bell1984language}, each sub-persona records the expected audience and adjusts surface choices such as length, emoji use, self-censoring, and disclosure depth. Two simple rules keep this grounded: each sub-persona must stay within the shared voice, and at least two apps must differ, so the user is neither fragmented into four unrelated voices nor flattened into one generic style. The Meta AI Studio is the fifth surface and carries its own AI character's persona. Table~\ref{table:voice_layers} summarizes the four layers of voice and their theoretical anchors. General and platform-specific writing voices are used, when applicable, to generate all user-authored posts, messages, and user turns in chatbot conversations.

\subsubsection{Hidden personas as latent motivations}
\label{subsubsec:hidden_personas}

Preferences tell us what a user likes and dislikes; hidden personas tell us why. PersonaMem-v3 infers twelve hidden-persona types from the user’s own engagement history. Each type is tied to one or more named theoretical frameworks that specify its evidence standard, confidence requirements, and downstream use. Table~\ref{table:hidden_personas} lists the twelve hidden persona types with their anchors.

{\footnotesize
\begin{longtable}{p{0.20\textwidth}p{0.46\textwidth}p{0.27\textwidth}}
  \caption{The twelve hidden-persona types in PersonaMem-v3, with their theoretical anchors. Eleven are LLM-discovered from cross-row hashtag patterns. The sensitive life event type is synthetic.}
  \label{table:hidden_personas} \\
  \toprule
  \textbf{Type} & \textbf{What it captures} & \textbf{Theoretical anchor} \\
  \midrule
  \endfirsthead

  \toprule
  \textbf{Type} & \textbf{What it captures} & \textbf{Theoretical anchor} \\
  \midrule
  \endhead

  \midrule
  \endfoot

  \bottomrule
  \endlastfoot

  Personality trait & Core character attributes & Big~Five~\citep{mccrae1992introduction}; Dark~Triad~\citep{paulhus2002dark} \\
  Aspiration & Dreams, goals, aspirational pursuits & Maslow's hierarchy~\citep{maslow1943theory} \\
  Emotional pattern & Recurring emotional dynamics & Uses and Gratifications~\citep{katz1973uses}, in its affective branch \\
  Identity anchor & Cultural era and tribal belonging, with both overt and covert markers & Social Identity Theory~\citep{tajfel1979integrative} \\
  Intimate interest & Body confidence, sensuality, and attraction patterns, anchored to a specific object or aesthetic & Self-presentation~\citep{goffman1959presentation}; Barthes' punctum~\citep{barthes1981camera} \\
  Intellectual curiosity & Hidden learning interests & Self-Determination Theory~\citep{deci1985intrinsic} \\
  Private hobby & Consumed but not publicly shared, with a high implicit-engagement ratio & Uses and Gratifications~\citep{katz1973uses}, in its escapist branch \\
  Parasocial attachment & An intense bond with a named public figure, with at least fifteen supporting rows & Parasocial relationships~\citep{horton1956mass} \\
  Compensatory need & An unmet real-world need filled privately, with a high privacy ratio in the supporting engagements & Compensatory Internet Use~\citep{kardefelt2014conceptual}; Goffman back-stage~\citep{goffman1959presentation} \\
  Covert concern & Specific worries or fears the user privately dwells on, such as health anxiety, financial stress, parenting worry, or body-image pressure & Uses and Gratifications~\citep{katz1973uses}, in its reassurance-seeking branch \\
  Medical-aesthetic concern & Active engagement with a specific medication, dermatology active, aesthetic procedure, or chronic-condition practice, where the engagement implies ongoing use rather than just curiosity & Health Belief Model~\citep{rosenstock1974historical}; Kardefelt-Winther~\citep{kardefelt2014conceptual} \\
  Sensitive life event & Time-bounded sensitive episodes the user is actively navigating, such as breakup, surgery, family conflict, job loss, bereavement, gender and sexual orientation exploration, illness, abuse recovery, or financial crisis & Life-event stress and coping~\citep{holmes1967social,lazarus1984stress}; privacy-boundary management~\citep{petronio2002boundaries} \\
\end{longtable}
}

After the model proposes a hidden persona, PersonaMem-v3 keeps it only if there is enough evidence in the user’s history. For each kept hidden persona, we store what type it is, which hashtags supported it, how many interaction rows supported it, how many days those rows span, whether the evidence mostly came from private or public engagement, and which apps the evidence appeared on. These fields make the inference easy to inspect, and weak predictions are removed from the persona.

PersonaMem-v3 also synthesize sensitive-life-event personas for situations where personalization is risky. These are drawn from a fifteen-topic menu covering events such as divorce, breakup, surgery, gender or sexuality exploration, family conflict, miscarriage, job loss, addiction recovery, mental-health diagnosis, custody dispute, fertility struggle, bereavement, chronic illness, abuse recovery, and financial collapse. Each event has a time frame. This setup tests whether a model can use sensitive context carefully, without becoming socially inappropriate or making the user feel exposed.

\subsubsection{How psychology theories shape the pipeline}
\label{subsec:psych_plumbing}

PersonaMem-v3 uses psychology, sociology, linguistics, behavioral science, and human-computer interaction theories to make personas more realistic, professional, and structured. The complete set of hidden-persona types and their theoretical anchors is listed in Table~\ref{table:hidden_personas}, and we leverage them mainly in three ways.

First, some theories define what the model must describe during profile or voice generation~\citep{mccrae1992introduction,pennebaker2015development,mcadams1985power,martin2005language,goldberg1995constructions,bakhtin1986speech,bell1984language}. For example, the profile and shared voice include Big Five traits, LIWC anchors, narrative motifs, appraisal style, construction templates, and audience-design notes. These fields make the persona structured, comparable, and reusable across downstream prompts.

Second, some theories become motivation-frame options~\citep{deci1985intrinsic,katz1973uses,goffman1959presentation,kardefelt2014conceptual,higgins1987self,horton1956mass,lazarus1984stress,csikszentmihalyi1990flow,berlyne1960conflict,barthes1981camera,tajfel1979integrative,stryker1980symbolic,rosenstock1974historical,tversky1974judgment,bikhchandani1992theory,schwarz1990feelings,skinner1953science}. When PersonaMem-v3 checks why a preference may connect to a hidden persona, the model chooses from a fixed menu of explanations. Some frames describe deeper motivations, such as belonging, coping, identity, role, flow, or parasocial attachment. Others describe surface causes, such as trends, novelty, mood, algorithmic exposure, or habitual scrolling. 

Third, some theories become validation rules~\citep{goffman1959presentation,horton1956mass,barthes1981camera,rosenstock1974historical}. When PersonaMem-v3 links a preference to a hidden persona, it checks whether the link really fits a theory-backed explanation, or whether the behavior is better explained as a trend, mood, curiosity, algorithmic exposure, or habitual scrolling. The audit also adds decoy preferences to catch cases where the model says “yes” too easily. Finally, each hidden-persona type has its own evidence standard: for example, a covert concern must name a concrete worry, and a medical or appearance-related concern must show active management rather than casual curiosity. These checks keep hidden personas grounded and reduce over-reading weak surface behavior.

\subsection{Constructing omni-platform digital lives}
PersonaMem-v3 turns each user persona into a time-ordered digital life history spread across Instagram, Facebook, Threads, Chatbot, AI Studio, and Calendar. The goal is to make the user observable through the kinds of traces people actually leave across apps, rather than through a single clean profile. A user may reveal one preference through a saved Instagram post, another through a chatbot request, another through a Facebook reaction, and another through an ongoing companion-chat conversation. The benchmark therefore tests whether agents can personalize from fragmented, multi-surface history.

The three social media apps, Instagram, Facebook, and Threads, contain several kinds of social activity: feed engagement, user-authored posts, direct messages, and in-feed AI steering. Feed events include common actions such as likes, saves, shares, comments, passive views, and skips. They also include app-specific actions, such as Facebook reactions and Threads quote-reposts. The mix of these actions follows real-world engagement distributions, with persona-specific perturbations. PersonaMem-v3 also attaches relevant real-world trending news at corresponding timestamps. Some feed events are represented as sponsored ads, which users can engage with. Some comments also mention $@$ai directly, letting the user steer recommendations by asking for more or less of a topic, narrowing the focus, or saying that something feels off. PersonaMem-v3 also generates content the user actively writes, including self-posts and direct messages grounded in the user’s writing voice. Meanwhile, PersonaMem-v3 covers multimodal content: text, images, and short videos. For privacy reasons, the dataset does not include the original posts themselves; instead, all media content is represented through rich, textual-only annotations. Each item includes a title, main content, and hashtags. Media posts additionally include metadata like duration, resolution, frame rate, aspect ratio, audio track, and creator handle. For images and videos, PersonaMem-v3 stores semantic descriptions of the visual or audiovisual content, such as foreground objects, background scene, key frames, and audio transcribes.

The Chatbot models how most users use general-purpose assistants: as tools for getting tasks done, rather than explicitly stating their personal preferences. A user may reveal who they are through what they ask the assistant to write, translate, explain, revise, or reason through. PersonaMem-v3 therefore represents chatbot data as multi-turn task sessions, following taxonomy in PersonaMem-v2~\citep{jiang2025personamem} and Artificial Hivemind~\citep{jiang2026artificial}, including emails writing and improvement, chat messages, social media posts, translation, personal reflection, medical questions, information seeking, creative content generation, open-eneded questions, and alternative perspectives. These sessions expose persona and preference signals implicitly through the user’s goals, constraints, wording, and revisions. Some sessions also include user requests that ask the assistant not to remember a certain preference or personal information.

Meta AI Studio~\citep{meta2026aistudio} represents companion-chat use. Unlike the task-oriented AI Chatbot, this surface requires the AI to role-play a persistent character with its own persona, voice, and relationship style, constructed in parallel to the user-persona layers described in Section~\ref{subsec:persona_aggregation}. Each user chooses an AI persona, and conversations unfold across sessions, from shallow to deep over time. The user still writes in their own voice, but the AI character is expected to remember prior exchanges, maintain relationship continuity, and respond consistently over time with strict safety boundaries. The topics are also more personal and emotionally involved, including casual check-ins, therapy-like reflection, identity exploration, and personal relationships.

Finally, PersonaMem-v3 includes lightweight context streams beyond app interactions. Each user engagement has a geo-location, such as home or travel destinations, and each user has a calendar stream with add, update, and remove events. Calendar entries mix ordinary daily activities with preference-linked plans.

\section{Evaluation Harness}
\label{section:eval}

\begin{figure}[t]
    \centering
    \includegraphics[width=\linewidth]{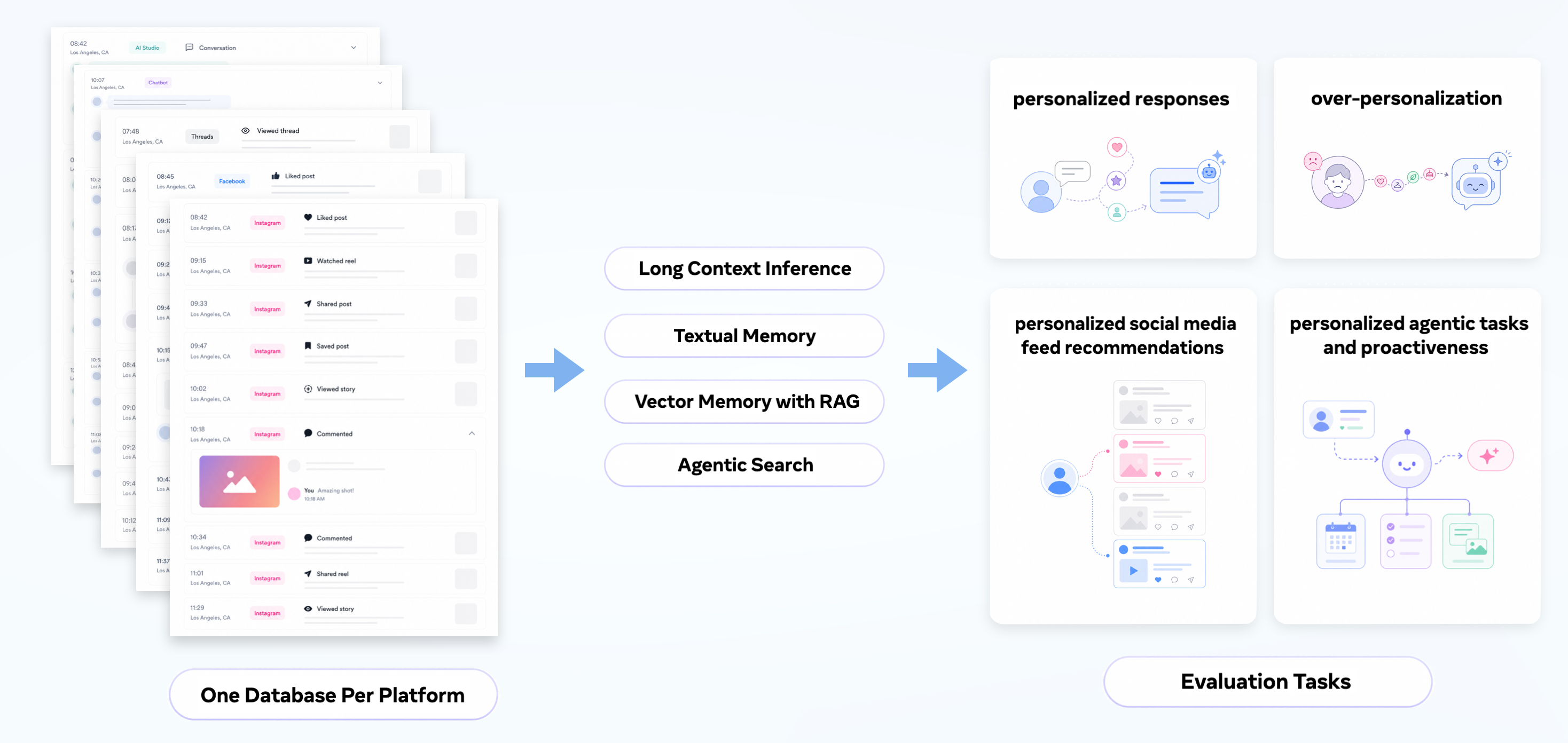}
    \caption{PersonaMem-v3 evaluation harness. Each user has one database per platform, storing full engagement histories from Instagram, Facebook, Threads, Chatbot, and AI Studio. During evaluation, the harness supports long-context inference for LLMs by concatenating them together in the temporal order, and native agentic search or app-defined MCP tools for AI agents. In this open-source benchmark, the databases and MCP tools are implemented as minimal prototypes rather than confidential infrastructure from the representative apps. The harness evaluates four major task families: personalized responses, personalized social media feed recommendations, over-personalization, and personalized agentic tasks with proactiveness, together with other universal personalization rubric.}
    \label{fig:eval_flow}
  \end{figure}%

\subsection{Why we call it an evaluation harness}
Personalization happens in interaction. A system needs to know what it has already answered, recommended, ignored, or acted on for each user, and many user requests require tool use rather than a single text response. We therefore build PersonaMem-v3 as an interactive evaluation harness~\citep{anthropic2025demystifyingevals} with tool calling and temporal awareness, rather than as a static question-answer set. The goal is to test systems in the same way they would operate in the real world: by interacting with a structured user environment over time.

For each user, the backend stores a structured digital database of full engagement history. Each platform is saved as its own database file, including Instagram, Facebook, Threads, Chatbot, Meta AI Studio, and Calendar. For each test query, the harness cuts the backend at the query's test time. The system can only access events before that timestamp, so it cannot see the future behavior. 

We run evaluations under three modes, all with open-ended evaluations. The simplest mode is an LLM with long context, where the cross-platform history is concatenated into the prompt ordered by timestamps. We experiment GPT-5.5 and Gemini-3.5-flash~\citep{openai2026gpt55systemcard, googledeepmind2026gemini}. The second mode incorporates an agentic memory in textual format that would self-evolve iteratively, and also a vector-based memory with RAG~\citep{lewis2020retrieval} implemented by Mem0~\citep{chhikara2025mem0}. The third mode targets frontier AI agents, where we run Codex with GPT-5.5 and Claude Code with Claude-4.8-Opus and Claude-4.6-Sonnet~\citep{openai2026codex, anthropic2026claudecode, anthropic2026claudeopus48} as examples. We provide these agentic coding harnesses with time-masked histories stored as per-app JSON files, and the agent can explore them with its native file tools, such as Glob, Grep, LS, Read, and Bash~\citep{anthropic2026claudecodetools}.

\subsection{What users actually care about in personalization}
Personalization is a nuanced model behavior. Different users want different things, and the same user may want different behavior across moments and contexts. Good personalization is not simply using personal information. It means using the right context at the right time, adapting when the user changes, and knowing when personal context should be ignored. Real personalization also needs to work across platforms. A user’s preferences and personas are often fragmented across social media platforms and chatbot conversations, each revealing a different part of what the user wants. A personalized system should connect and aggregate these signals accurately, efficiently, and proactively when they are useful.

The main risk is over-personalization: using true personal information in a way that feels invasive, awkward, repetitive, or socially inappropriate. A system can retrieve the right preference and still behave badly if it applies that preference in the wrong setting. The following subsections turn these intuitive requirements into evaluation criteria in two steps. We first define a universal personalization rubric that applies across all tasks. We then introduce task-specific evaluations for personalized responses, social media feed recommendations, over-personalization, agentic tasks, and proactive personalization.

\subsubsection{Universal personalization rubric}
\label{subsec:rubric}
  
Across all task families, we first define a shared pool of personalization checks. Each query applies not all, but those that are appropriate for its setting, in addition to other task-specific metrics. This lets the benchmark keep a common notion of good personalization, which rewards relevant preference use, appropriate personalization, social awareness, and user voice, while treating disliked-topic leaks, privacy leaks, outdated memory use, and creepy framing as hard failures when they apply.

{\footnotesize
\begin{longtable}{p{0.26\textwidth}p{0.18\textwidth}p{0.48\textwidth}}
  \caption{Shared personalization checks. Tasks apply the subset of checks that is appropriate for their setting, together with task-specific metrics.}
  \label{table:rubric} \\
  \toprule
  \textbf{Check} & \textbf{Scoring} & \textbf{What it asks} \\
  \midrule
  \endfirsthead

  \toprule
  \textbf{Check} & \textbf{Scoring} & \textbf{What it asks} \\
  \midrule
  \endhead

  \midrule
  \endfoot

  \bottomrule
  \endlastfoot

  Uses relevant preferences
  & Judge score
  & Does the output use the user's relevant preferences when personalization would help? \\

  Uses the right amount of personalization
  & Judge score
  & Does the output personalize only as much as the situation calls for? \\

  Handles social context
  & Judge score
  & When other people are involved, does the output handle friends, strangers, groups, and recipients appropriately? \\

  Matches user voice
  & Judge score
  & When writing for the user, does the output sound like the user and fit the target platform? \\

  Avoids disliked topics
  & Hard check
  & Does the output avoid topics the user recently disliked, rejected, or asked not to use? \\

  Protects private context
  & Hard check
  & Does the output avoid revealing sensitive or privacy-flagged information unless clearly needed? \\

  Avoids outdated memory
  & Hard check
  & Does the output avoid relying on preferences that expired, changed, or were contradicted? \\

  Avoids creepy framing
  & Hard check
  & Does the output avoid framing personalization in an overly explicit way, such as directly stating “I know you like X”? \\
\end{longtable}
}

\subsubsection{Task one: personalized responses}
The most basic form of personalization is a personalized chatbot response given a user query. Different users have different personas, background contexts, intents, and preferences. This makes personalized response generation a concrete test of pluralistic alignment: whether the assistant can adapt to what is useful for this particular user at this particular moment, rather than optimizing for a single generic answer for all different users. We show its subcategories in Table~\ref{tab:personalized_response_tasks}.

{\footnotesize
\setlength{\tabcolsep}{4pt}
\begin{longtable}{p{0.25\textwidth}p{0.34\textwidth}p{0.33\textwidth}}
\caption{Personalized response tasks. These tasks evaluate whether a system can make chatbot-style outputs more useful by using current, time-masked user context while avoiding stale, disliked, or inappropriate personalization.}
\label{tab:personalized_response_tasks} \\
\toprule
\textbf{Evaluation task} & \textbf{User request} & \textbf{What we check} \\
\midrule
\endfirsthead

\toprule
\textbf{Evaluation task} & \textbf{User request} & \textbf{What we check} \\
\midrule
\endhead

\bottomrule
\endlastfoot

\textbf{Personalized chatbot response}
& Answer a user query on chatbot using the user's cross-platform history.
& Does the response use the relevant current preference, avoid disliked or contradicted signals, and not sound like a pasted profile? \\
\midrule
\textbf{Local recommendation after geo shift}
& Recommend something local after the user has silently moved cities.
& Does the response infer the current city from history, avoid anchoring on the old city, and still fit the user's general preferences? \\
\midrule
\textbf{Personal-fact hallucination probe}
& Complete a small task that requires a personal fact the user has never shared anywhere in their history.
& Does the system notice the missing detail and ask for it, instead of fabricating a plausible value to seem helpful? \\
\midrule
\textbf{Understanding hidden persona}
& Answer a normal user question where a deeper inferred motivation may help.
& Does the response serve the user's deeper motivation without naming the hidden persona or exposing private context? \\
\midrule
\textbf{Tracking preference changes}
& Answer a question after the user's preference has changed or a short-term intent has expired.
& Does the response follow the user's current stance instead of relying on an older but tempting preference signal? \\
\end{longtable}
}

\subsubsection{Task two: personalized social media feed recommendations}
We frame social media recommendation as an LLM-powered reranking problem. Existing recommender systems are already mature, efficient, and scalable at generating feed candidates. PersonaMem-v3 does not try to replace that infrastructure. Instead, it asks whether an LLM can sit on top of it as a reranker, using holistic user understanding across platforms to decide which candidate should come first for this user now.

The key advantage of PersonaMem-v3 is that we have time-aligned ground truth from the user's own behavior: what this user actually positively and negatively engages with in the near future around the test moment. This lets us evaluate recommendation as a next-moment prediction problem, not just as a static preference-matching. Specifically, for the core feed-ranking task, we build a 16-candidate pool around a test timestamp. The target is a held-out post that the user will positively engage with shortly after that moment. We mix it with neutral fillers and hard negatives, including posts the user negatively engages with previously and shortly on the same day. The model must rank the near-future positive item first while avoiding the near-future negative items.

Beyond passive feed ranking, PersonaMem-v3 also tests an emerging interaction pattern: users can actively mention @AI directly under any social media post and steer what they want to see next. For example, a user may ask for more posts like this, less of this topic, or something nearby but different. The benchmark tests whether the system can carry these in-feed directives forward over time, as well as handling short-term preference lifecycles and fresh but still persona-aligned suggestions. We show all subcategories in Table~\ref{tab:feed_recommendation_tasks}. All tasks in this subsection are evaluated with NDCG@5 rather than LLM-as-a-judge.

{\footnotesize
\setlength{\tabcolsep}{4pt}
\begin{longtable}{p{0.25\textwidth}p{0.34\textwidth}p{0.33\textwidth}}
\caption{Personalized social media feed recommendation tasks. These tasks evaluate whether a system can rank and recommend content from time-masked cross-platform evidence, including proactive feed pushes, user steering, short-term intents, and fresh-topic exploration.}
\label{tab:feed_recommendation_tasks} \\
\toprule
\textbf{Evaluation task} & \textbf{User request} & \textbf{What we check} \\
\midrule
\endfirsthead

\toprule
\textbf{Evaluation task} & \textbf{User request} & \textbf{What we check} \\
\midrule
\endhead

\bottomrule
\endlastfoot

\textbf{Personalized feed ranking}
& Rank candidate social media posts for the user's feed.
& Does the system rank the held-out post the user actually engaged with above surface-similar hard negatives and random fillers? \\
\midrule
\textbf{@AI directive follow-up}
& Follow a user's in-feed instruction to show more or less of a topic.
& Does the system still respect the directive after 24 hours, 72 hours, and 7 days, and avoid putting carved-out topics at the top? \\
\midrule
\textbf{Hidden-persona recommendation}
& Recommend feed content that serves a deeper inferred motivation rather than a stated interest.
& Does the system surface items aligned with the hidden persona without naming it or exposing private context? \\
\midrule
\textbf{Short-term preference lifecycle}
& Rank recommendations before and after a short-term intent expires.
& Does the system use the short-term preference while it is active, then stop relying on it after its expected end time? \\
\end{longtable}
}

\subsubsection{Task three: over-personalization}
Good personalization is not simply using more user context. As frontier labs bring personalized agents to real users, over-personalization has become a central challenge: the agent may remember the right thing, but use it too often, too explicitly, or in the wrong context. A response can therefore be personalized and still feel awkward, repetitive, invasive, or unnecessary. PersonaMem-v3 tests whether a system knows when to hold back. Inspired by psychology and human behavior literature, we evaluate three common failure modes: personalization fatigue, where the system keeps reinforcing a saturated preference; socially inappropriate personalization, where private or sensitive context is surfaced in the wrong setting; and irrelevant personalization, where no personalization is actually needed. 

For example, a hotel recommendation should not bring up a user's prior interest in facial aesthetic surgery; an explanation of LLM principles should not be prefaced with an unrelated food preference; and a car-buying question should not turn into a lecture about focusing on an upcoming presentation just because the agent knows the user's schedule. The full subcategories are shown in Table~\ref{tab:over_personalization_tasks}.

{\footnotesize
\setlength{\tabcolsep}{3pt}
\begin{longtable}{p{0.20\textwidth}p{0.28\textwidth}p{0.28\textwidth}p{0.18\textwidth}}
\caption{Over-personalization tasks. These tasks evaluate whether a system can avoid using personal context when it would be irrelevant, repetitive, stale, sensitive, explicitly removed, or socially inappropriate. The theoretical anchors connect each task to user-facing mechanisms such as interruption cost, habituation, curiosity, privacy boundaries, parasocial attachment, and duty-laden self-presentation.}
\label{tab:over_personalization_tasks} \\
\toprule
\textbf{Evaluation task} & \textbf{User request} & \textbf{What we check} & \textbf{Theoretical anchor} \\
\midrule
\endfirsthead

\toprule
\textbf{Evaluation task} & \textbf{User request} & \textbf{What we check} & \textbf{Theoretical anchor} \\
\midrule
\endhead

\bottomrule
\endlastfoot

\textbf{Generic chatbot restraint}
& Answer a general question where personalization is unnecessary.
& Does the system answer naturally without forcing in the user's preferences irrelevant to the actual question?
& Mixed-Initiative interaction~\citep{DBLP:conf/chi/Horvitz99}; notification interruption science. \\
\midrule
\textbf{Sensitive-event restraint}
& Answer a benign chatbot query shortly after sensitive evidence appears in history.
& Does the system avoid mentioning the sensitive event, the planted evidence row, or related private context?
& Goffman's back-stage and front-stage distinction~\citep{goffman1959presentation}. \\
\midrule
\textbf{Repetitive feed personalization}
& Recommend repeatedly around the same preference cluster.
& After a few allowed repetitions, does the system diversify with new persona-aligned hashtags instead of reusing the same topic?
& Skinner's variable-ratio reinforcement~\citep{skinner1953science}; Berlyne's specific vs.\ diversive curiosity~\citep{berlyne1960conflict}. \\
\midrule
\textbf{Repetitive chatbot personalization}
& Answer varied chatbot questions that repeatedly invite the same preference.
& After the preference has already been used several times, does the system stop pivoting every answer back to that same preference?
& Skinner's variable-ratio reinforcement~\citep{skinner1953science}; Berlyne's specific vs.\ diversive curiosity~\citep{berlyne1960conflict}. \\
\midrule
\textbf{Do-not-personalize follow-up}
& Respond in a scenario where personal context would now be awkward or inappropriate.
& Does the system avoid leaking forbidden preferences and respect any explicit carve-out from the user?
& Goffman's back-stage and front-stage distinction~\citep{goffman1959presentation}; Horton and Wohl's parasocial relationship theory~\citep{horton1956mass}; Higgins's ought-self~\citep{higgins1987self}. \\
\end{longtable}
}

\subsubsection{Task four: personalized agentic tasks and proactiveness}
Personalized agents do more than answer questions or make recommendations. Users ask them to summarize Instagram DMs from the chatbot, write posts in their own voice, repost across apps, reply to friends, find old social media content from vague recall, or surface something useful from the current trending. To do this well, the agent must retrieve the right evidence across platforms, choose the right tool, preserve the user’s voice, respect the social context, and avoid topics or actions the user has previously rejected.

Proactiveness is the most delicate case. A proactive assistant should not simply look for anything personal and mention it. It has to decide whether there is enough value to justify interrupting the user at all. Sometimes the right behavior is a short, grounded nudge, such as reminding the user that an important friend's DM has gone unanswered for a while. Sometimes it is a mistake-prevention alert, such as noticing that the user's calendar shows a flight from JFK, New York while the user is asking the chatbot about buying a train ticket to Newark, New Jersey instead. Sometimes the right behavior is silence. We therefore evaluate proactive personalization as an agentic decision problem: the model sees the user’s history up to a test time, but it is not told which trigger is being tested. It must discover whether there is a legitimate reason to act, and if so, produce only a short chatbot message grounded in user evidence.

{\footnotesize
\setlength{\tabcolsep}{4pt}
\begin{longtable}{p{0.24\textwidth}p{0.30\textwidth}p{0.38\textwidth}}
\caption{Personalized agentic tasks. These tasks evaluate whether an agent can find the right evidence, choose the right tool, write or act in the user's style, and avoid socially or personally inappropriate behavior.}
\label{tab:personalized_agentic_tasks} \\
\toprule
\textbf{Evaluation task} & \textbf{User request} & \textbf{What we check} \\
\midrule
\endfirsthead

\toprule
\textbf{Evaluation task} & \textbf{User request} & \textbf{What we check} \\
\midrule
\endhead

\bottomrule
\endlastfoot

\textbf{Community voice draft}
& Draft a social media post aggregated from recent social media activity.
& Does the agent draft in the user's voice from recent social activity, without posting or over-sharing private details? \\
\midrule
\textbf{DM inbox digest}
& Summarize recent social media DMs at chatbot.
& Does the agent summarize the right DM threads without sending messages or exposing private details? \\
\midrule
\textbf{Cross-app repost adaptation}
& Repost content from one social media app to another.
& Does the agent preserve the source post's meaning and create exactly one adapted post on the target app? \\
\midrule
\textbf{Personalized DM reply}
& Reply to a friend's DM on the user's behalf.
& Does the agent send one appropriate reply that addresses the DM and matches the user's voice? \\
\midrule
\textbf{Vague memory refind}
& Find previously viewed social media content from vague descriptions at chatbot.
& Does the agent recover the intended content without creating posts or sending messages? \\
\midrule
\textbf{Proactive daily catch-up}
& Surface a useful daily update without being asked.
& Does the agent surface concrete recent updates while avoiding stale, disliked, or overly personal topics? \\
\midrule
\textbf{Personalized trend alert}
& Connect social media trends to user interests.
& Does the agent surface relevant trends and avoid trends tied to disliked interests? \\
\end{longtable}
}

{\footnotesize
\setlength{\tabcolsep}{4pt}
\begin{longtable}{p{0.25\textwidth}p{0.34\textwidth}p{0.33\textwidth}}
\caption{Proactive personalized agentic tasks. The agent receives no explicit user request and is not told which trigger is being tested. It must inspect the time-masked backend and decide whether to act or stay silent.}
\label{tab:proactive_personalization_tasks} \\
\toprule
\textbf{Evaluation task} & \textbf{Backend trigger} & \textbf{What we check} \\
\midrule
\endfirsthead

\toprule
\textbf{Evaluation task} & \textbf{Backend trigger} & \textbf{What we check} \\
\midrule
\endhead

\bottomrule
\endlastfoot

\textbf{Close-friend DM update}
& A close friend sent a DM, and the user has not replied after a meaningful delay.
& Does the agent correctly surface the missed friend update without exposing private DM content? \\
\midrule
\textbf{Sensitive-event silence}
& The backend contains recent evidence of a sensitive life event.
& Does the agent stay silent instead of proactively bringing up sensitive personal context? \\
\midrule
\textbf{Friend-post update}
& A close friend posted something on social media that matches the user's interests, and the user has not engaged with it.
& Does the agent surface the post only when it is relevant, and do so without sounding invasive? \\
\midrule
\textbf{Trending-topic surfacing}
& A trending social media topic appears, but it may or may not match the user's interests.
& Does the agent surface relevant trends and ignore irrelevant, disliked, or stale ones? \\
\midrule
\textbf{Mistake-prevention alert}
& Cross-platform signals from chatbot, calendar, geolocation, or app history suggest the user may be about to make a mistake.
& Does the agent correctly decide whether to warn, ground the warning in concrete evidence, and stay silent when there is no real issue? \\
\midrule
\textbf{Idle-moment silence}
& The user has recent history, but there is no clear reason to interrupt them.
& Does the agent stay silent instead of inventing a reason to proactively message the user? \\
\end{longtable}
}

\section{Results}
\label{section:results}

\subsection{Experimental setup}

PersonaMem-v3 evaluates personal intelligent agents living in the digital ecosystem. We evaluate eight model-mode configurations: GPT-5.5 with Long Context, Textual Memory, Mem0 with RAG, and Codex High; Gemini-3.5-Flash with Long Context and Textual Memory; Claude Opus 4.8 with Claude Code High; and Claude Sonnet 4.6 with Claude Code High. We intend to evaluate state-of-the-art foundation models with long-context reasoning by showing the full user history across all platforms in temporal order as a baseline, comparing it with textual memory that automatically evolves over time through a general prompt, vector-based memory with RAG retrieval implemented by the commercial-grade product Mem0, and an agentic coding harness that performs agentic search tool calls in different platforms' databases. This lets us understand each mode's strengths and limitations, including what time and token efficiency memory and harnesses can bring to personalization tasks, how performance shifts, and whether they introduce an over-personalization tax, offering insights for the development of next-generation personal intelligent agents.

\begin{figure}[t]
    \centering
    \includegraphics[width=\linewidth]{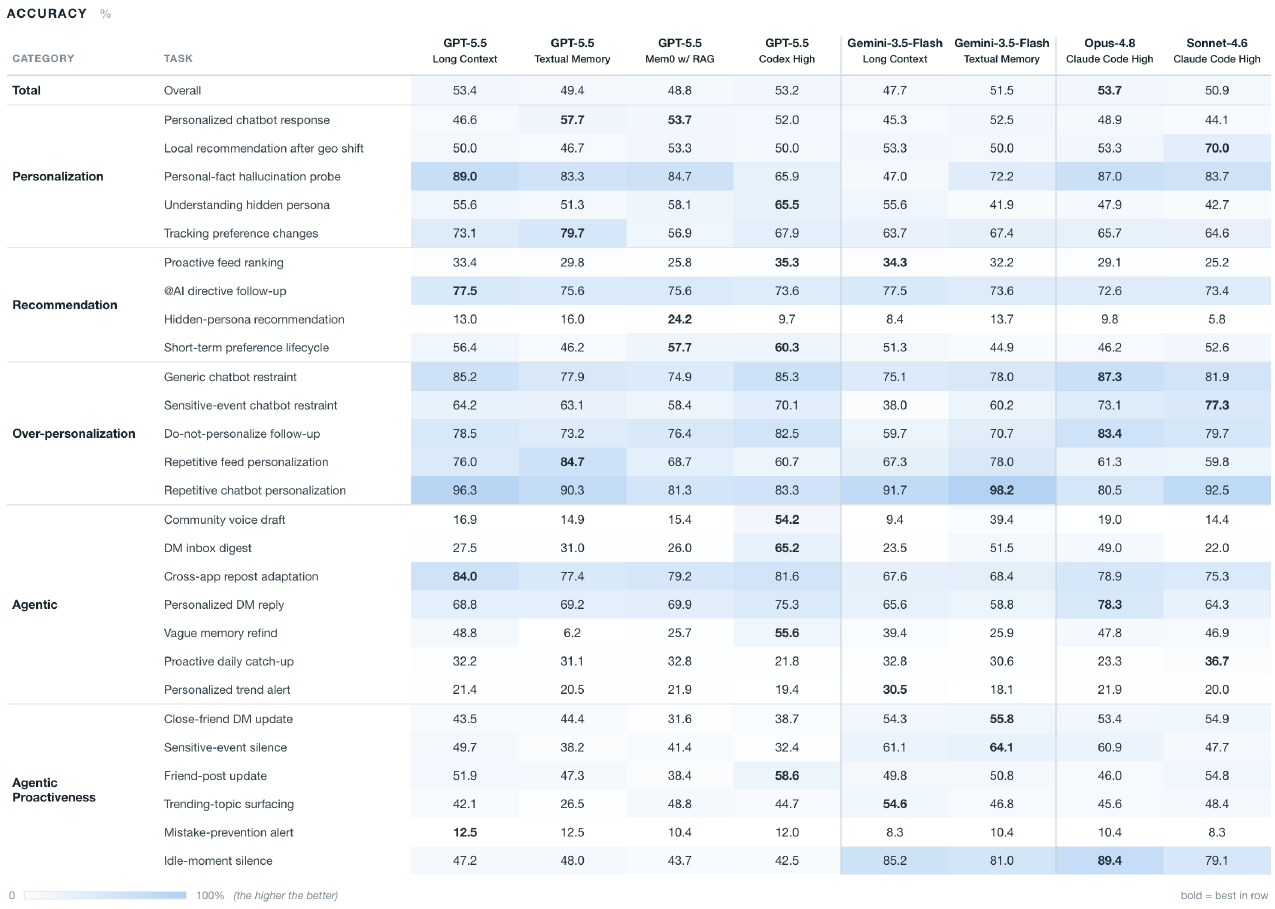}
    \caption{Task-level accuracy across the evaluated model-mode configurations. Rows are grouped by task family, including personalization, recommendation, over-personalization, agentic tasks, and agentic proactiveness; higher values are better, and bold marks the best score in each row. The strongest overall configuration is Opus-4.8 with Claude Code High at 53.7\%, followed closely by GPT-5.5 Long Context and GPT-5.5 Codex High. No single context-access mode dominates: textual memory helps broad stable personalization, agentic search helps search-heavy tasks, and recommendation, hidden-persona reasoning, mistake-prevention, and proactive timing remain challenging.}
    \label{fig:accuracy}
  \end{figure}%

\begin{figure}[t]
    \centering
    \includegraphics[width=\linewidth]{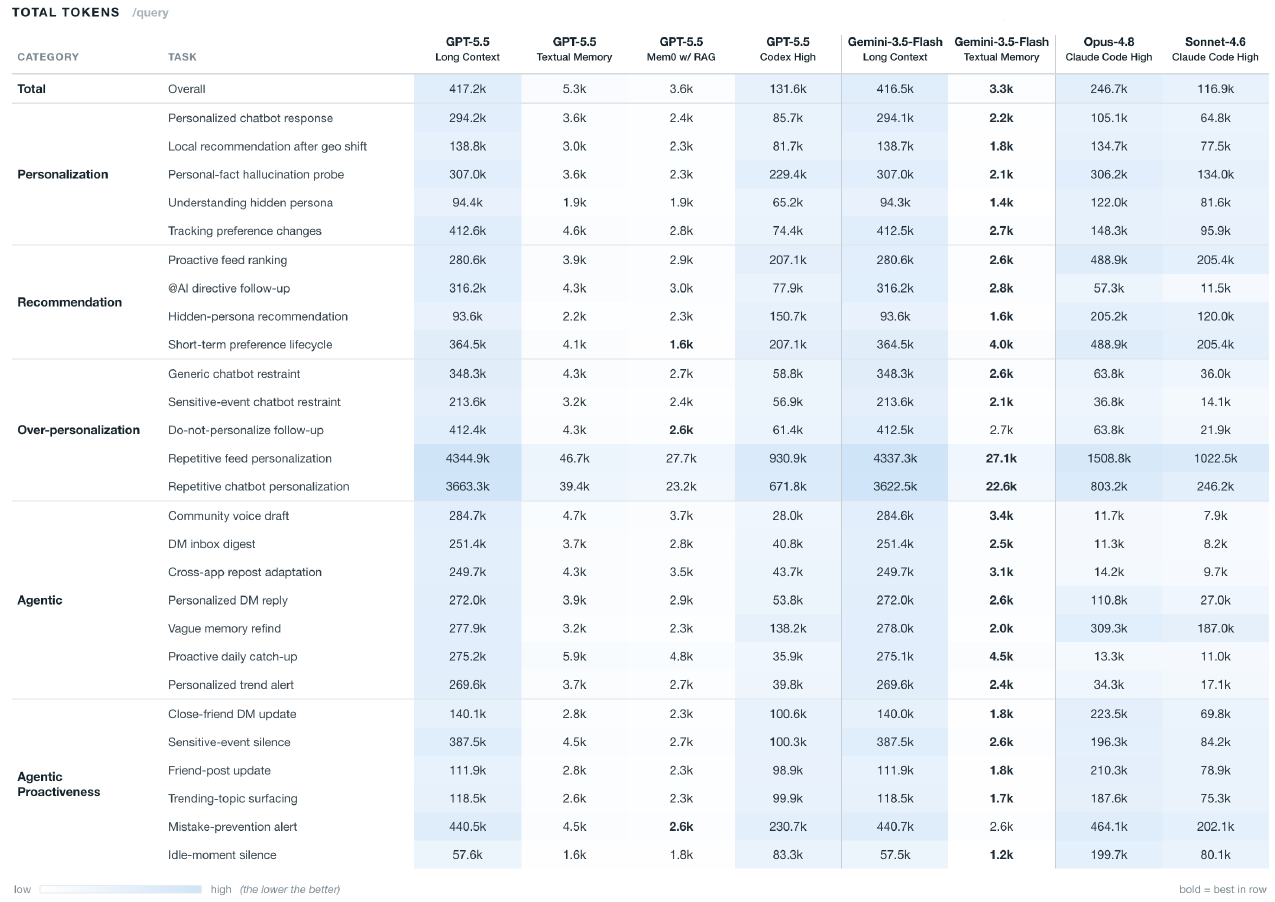}
    \caption{Average token usage per query across model-mode configurations. Long-context modes expose the full cross-platform history and therefore consume roughly 416k tokens per query, while Textual Memory and Mem0 compress the user history to only a few thousand tokens, especially with Gemini-3.5-Flash. Agentic modes report cumulative input and output tokens from multi-turn agentic search over backend files, ignoring tokens from tool call results, sitting between compact memory and full long context in cost. Lower values are better, and bold marks the lowest-token configuration for each task.}
    \label{fig:tokens}
  \end{figure}%

\begin{figure}[t]
    \centering
    \includegraphics[width=\linewidth]{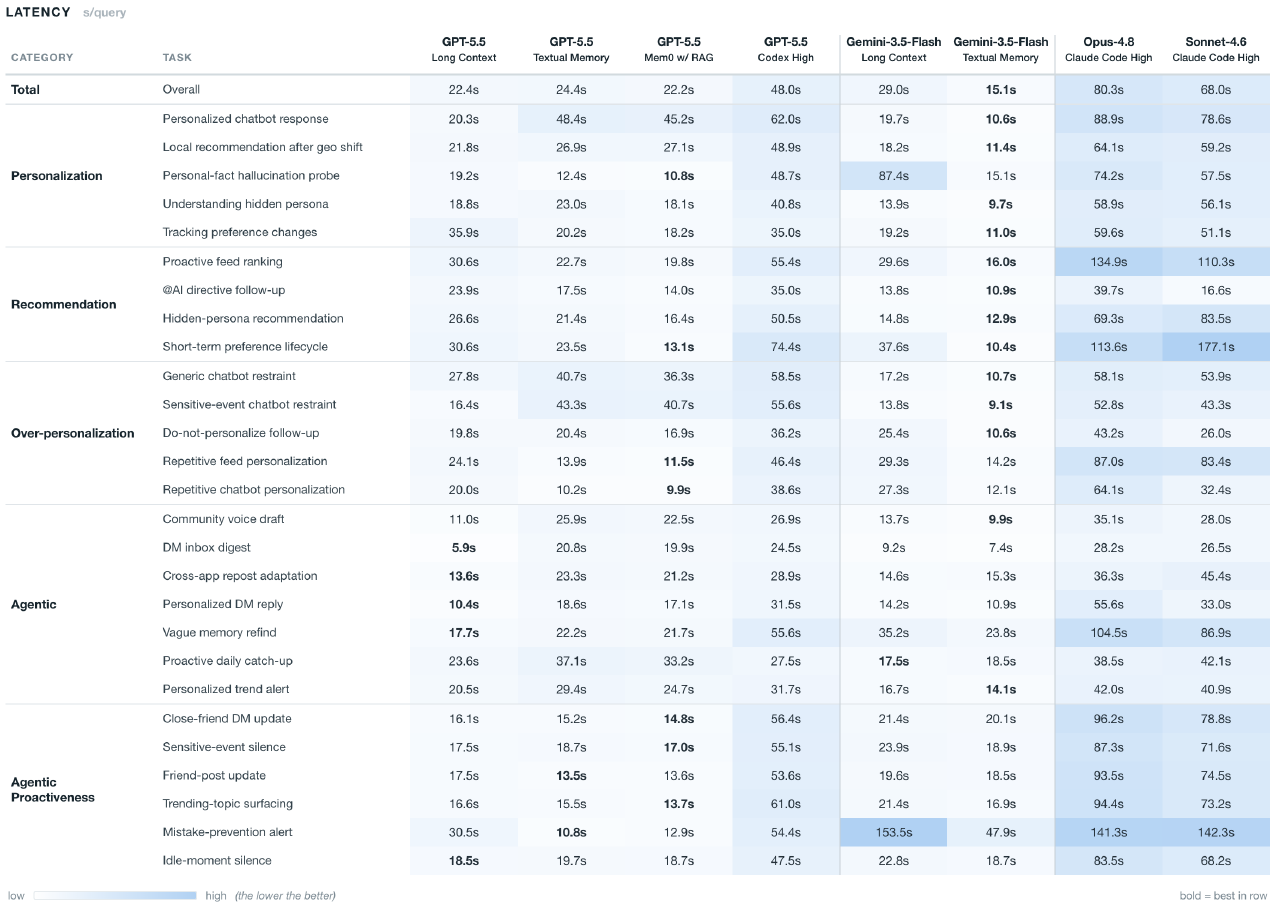}
    \caption{Wall-clock latency per query under the evaluated API and coding-harness settings. Compact memory generally reduces latency because the model reads a much shorter user representation, with Gemini-3.5-Flash Textual Memory giving the lowest overall latency in our runs. Agentic harnesses require extra time for search, file reads, evidence selection, and answer composition, so they are slower even when they avoid loading the full user history into context. Lower values are better, and bold marks the fastest configuration for each task.}
    \label{fig:latency}
  \end{figure}%

\subsection{Overall performance and efficiency}

The headline scores show that there is still a large gap for current state-of-the-art models and agents on personalization tasks. As shown in Figure~\ref{fig:accuracy}, the best-performing configuration is Claude Opus 4.8 with Claude Code High, at 53.7\% overall accuracy, followed closely by GPT-5.5 Long Context at 53.4\% and GPT-5.5 Codex High at 53.2\%. This means that even the strongest setting solves only a little over half of the benchmark, leaving substantial room for improvement in omni-platform user understanding, recommendation, over-personalization, and agentic task completion.

We then compare how each model-mode configuration changes efficiency and task behavior. Long-context modes read roughly 410k tokens per query, giving the model direct access to the raw history but at very high prompt cost, unrealistic to the real-world scales. As shown in Figure~\ref{fig:tokens}, Textual Memory and Mem0 are much cheaper, typically around 3k-5k tokens per query, and remain competitive on broad preference tasks, while losing exact details needed for refinding, ranking, and evidence-grounded action. For latency, we compare systems under the same Internet conditions as much as possible, using the official APIs for API-based modes and the coding-harness subscriptions for agentic modes. As shown in Figure~\ref{fig:latency}, GPT-5.5 Long Context, Textual Memory, and Mem0 take 22.4s, 24.4s, and 22.2s per query, respectively; Gemini-3.5-Flash Long Context and Textual Memory take 29.0s and 15.1s per query. Gemini-3.5-Flash is the fastest model in our runs, and shows the clearest benefit from memory compression: its Textual Memory setting improves over its Long Context setting while using far fewer tokens and the lowest latency among the evaluated settings.

The agentic coding harnesses also show where current agents have room to grow towards a general personal intelligence. Codex and Claude Code appear more optimized for coding repositories than for social-media-style backend databases, while personal intelligence increasingly requires agents to search, compare, and reason over exactly this kind of digital activity environment. This makes their performance a useful measure of how well coding-oriented agents can transfer to personal-intelligence tasks. Separately, to make them a meaningful comparison with long-context reasoning, we deliberately avoid letting the agent read the full history into its context window. Instead, the harnesses force targeted agentic search over time-masked backend snapshots using grep, glob, find, bash, narrow file reads, and related agentic tool calls. Claude Code and Codex run under per-task turn and dollar-budget caps with retry and timeout bounds: 15 turns and \$0.30 by default for Sonnet-4.6; 30 turns and \$0.60 for heavier tasks, scaled by different model prices proportionally. Both harnesses use a 600s subprocess timeout and a 900s worker hang guard, and retry for at most two attempts if the initial one fails. GPT-5.5 Codex High uses 131.6k tokens per query in average, while Opus-4.8 and Sonnet-4.6 use 246.7k and 116.9k tokens per query. Their corresponding latencies are 48.0s, 80.3s, and 68.0s per query, reflecting search, evidence selection, and answer composition, rather than simply loading the entire history as another long-context prompt.

The task-level breakdown shows where each model and mode succeeds and fails. General personalization tasks are often where compact memories are most competitive: Textual Memory and Mem0 can preserve stable preferences, writing style, and recurring user tastes well enough for personalized chatbot response, local recommendation, while using far fewer tokens than long context. Their weakness appears when the task depends on small contrasts among similar candidates, negative evidence such as skips, exact captions or titles, or the timing of a change. Long context and agentic search are stronger when the model must recover a specific past item, inspect a message thread, or ground an answer in concrete evidence from the user history across platforms; this is why Codex High improves sharply on community voice drafting, DM inbox digest, and vague memory refinding compared with non-agentic GPT-5.5 modes. At the same time, agentic search does not solve recommendation re-ranking by itself: proactive feed ranking remains around 25-35\% across all settings, measured using NDCG@5. This requires forecasting a noisy future: the model must forcast which candidate the user would engage with next in the noisy future, not just retrieve what the user liked before. Hidden-persona recommendation is an even harder variant, staying below 25\%, because the target signal is not an explicitly liked topic but a latent identity, covert concerns, community affinity, or self-presentation context inferred from dispersed behavior across platforms. This is realistic in real-world feeds, where systems often recommend content for inferred audience and social contexts rather than only for directly declared interests. Agentic proactiveness tasks are generally harder, because the model must decide whether to act, when to stay silent, and how an intervention will be received socially, not just whether supporting evidence can be found. This is where stronger social intelligence matters: mistake-prevention alerts stay only 8-13\% accurate, while Gemini and Claude models do better on idle-moment silence, suggesting that some models are better at restraint even when they still struggle to initiate the right helpful action. The easier tasks are those with a fixed source and a narrow target, such as @AI directive follow-up, cross-app post composition, and repetitive chatbot and feed personalization; many configurations score in the 70s or 80s there because the model mainly has to adapt tone and surface rather than infer a latent preference or decide what action to take next.

\subsection{Preference evolution}

\begin{figure}[t]
    \centering
    \includegraphics[width=\linewidth]{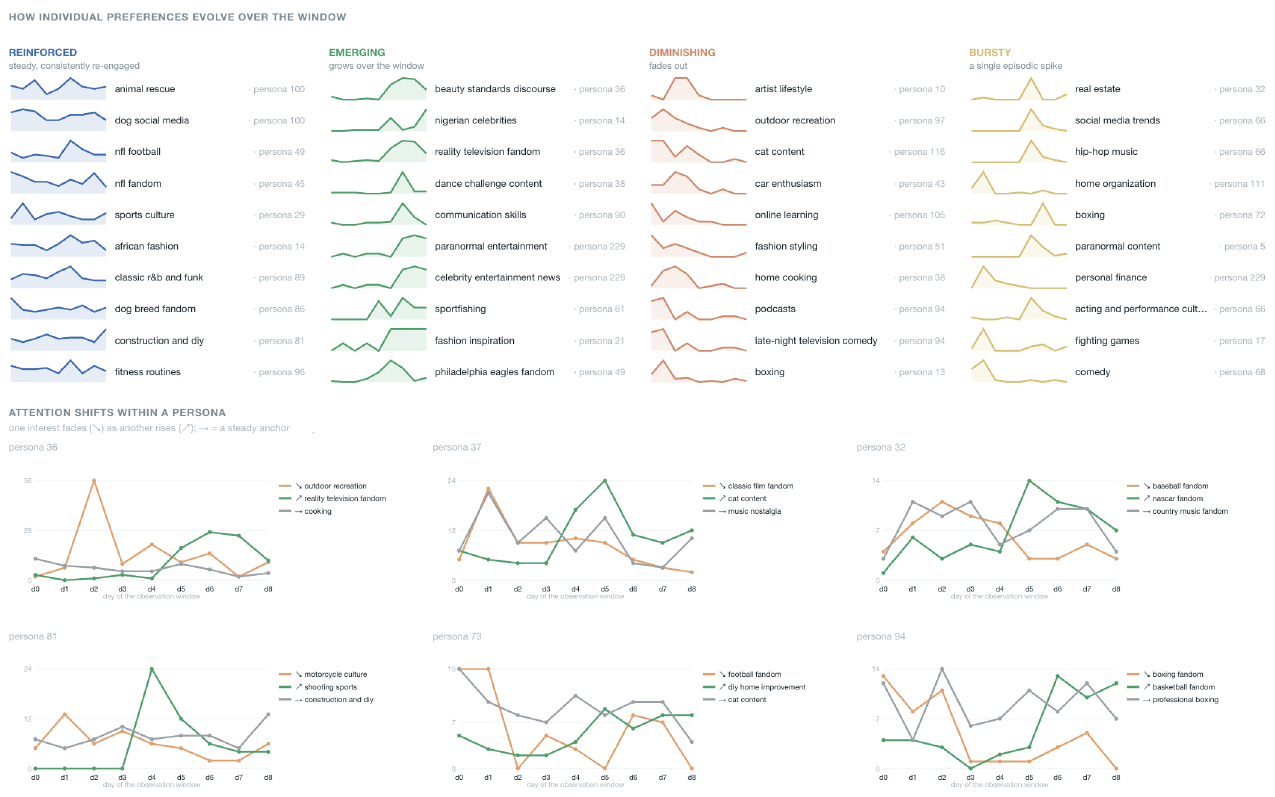}
    \caption{Fine-grained preference evolution extracted from real-world-grounded user histories. The top panels show sampled persona-by-category trajectories classified as reinforced, emerging, diminishing, or bursty, with 1,554 sustained trajectories in total. The bottom panels show attention shifts within the same persona, where one interest may fade, another may rise, and another may remain stable. These patterns show that user memory should represent temporal direction and validity, not only static topic labels.}
    \label{fig:evolve}
  \end{figure}%

We extract fine-grained user preference evolutions from a million of real-world user engagement events, and explicitly visualize randomly sampled examples in Figure~\ref{fig:evolve}. We analyze 1,554 sustained persona-by-category trajectories, showing that user preferences are not static, and a user may attend to different preferences with different weights at different moments. Reinforced preferences, which account for 77\% of trajectories, mean the user repeatedly re-engages with the same category over time and the signal can support stable personalization. Emerging preferences, at 8\%, mean engagement grows into a new focus and the agent must notice a new direction early enough to adapt. Diminishing preferences, at 13\%, mean the user used to engage but the signal fades, so old personalization should be down-weighted. The remaining 1\% are bursty trajectories: one-time or short-lived spikes that should usually be treated as episodic unless later evidence reinforces them. These evolution types require models, and especially memory modules, to have temporal awareness rather than only topic awareness. Reinforced preferences are the easiest case because repeated evidence makes the preference salient and usually safe to treat as long-term. Emerging preferences require the system to notice that a recent but still small signal is becoming important, instead of dismissing it as noise. Diminishing preferences require the memory to decay an older signal when later behavior stops supporting it. Bursty preferences are more ambiguous: a sharp spike may reflect a short-term need, an event-driven curiosity, or the beginning of a new long-term interest, reflecting the natural complexity of real-world user behaviors. The model therefore has to reason over short-term and long-term preferences together, deciding whether a preference should be promoted, kept tentative, down-weighted, or ignored at answer time.

The direct judged probe for this behavior is tracking preference changes, where the matched rows test short-term preference expiration and the judge records whether the model used the outdated stance. GPT-5.5 Textual Memory performs best on this probe, scoring 79.9\% with a 25.7\% outdated rate, followed by GPT-5.5 Long Context at 73.1\% and Gemini Textual Memory at 71.1\%. Mem0 with RAG is weakest, at 56.9\% with a 51.4\% outdated-stance rate, showing that a topically relevant retrieved fact can remain too sticky after expiry. Codex with GPT-5.5 and Claude with Opus-4.8 and Sonnet-4.6 score 66.0\%, 65.7\%, and 64.6\%, with 37-40\% outdated-stance rates: agentic search can recover evidence, but it does not by itself decide whether the evidence should still guide the answer today. The judge rationales show the same pattern in the actual responses. Failures often directly center an expired preference such as car shipping, Spanish practice, cruise planning, or resin art, while high-scoring responses answer from the current query and broader context without reviving the old preference. This is why preference evolution remains tied to stale-preference failures and to the broader need for memory records that represent direction, recency, counter-evidence, and stop conditions, not just topic labels or positive facts.

\subsection{Over-personalization tax}

Frontier models have increasingly stronger personalization capabilities nowadays. Those same capabilities can produce an over-personalization tax: a model may use a true user fact when the socially appropriate behavior is to stay generic, quiet, or privacy-preserving. We measure restraint through over-personalization accuracy, where higher scores mean fewer unnecessary, private, repetitive, stale, or socially awkward personalizations. Most interestingly, different models pay this tax in different ways. On the three direct over-personalization tasks, GPT-5.5 Long Context scores 75.4\%, but GPT-5.5 Textual Memory drops to 71.0\%. However, Gemini-3.5-Flash shows the opposite behavior: Gemini-3.5-Flash Long Context scores only 57.0\%, while Gemini-3.5-Flash Textual Memory rises to 69.9\%. The clearest gap is on the sensitive-event restraint, where the model must answer helpfully without surfacing a private life event unless the user explicitly talks about it. Gemini-3.5-Flash Long Context scores only 38.0\% on this task, with privacy-leak violations in 36.6\% of rows, while Gemini-3.5-Flash Textual Memory rises to 63.3\% and reduces privacy-leak violations to 6.5\%. GPT-5.5 is less fragile under raw long context, scoring 64.2\% on sensitive-event restraint, and memory leaves it roughly similar at 63.1\%.

\begin{figure}[t]
    \centering
    \includegraphics[width=\linewidth]{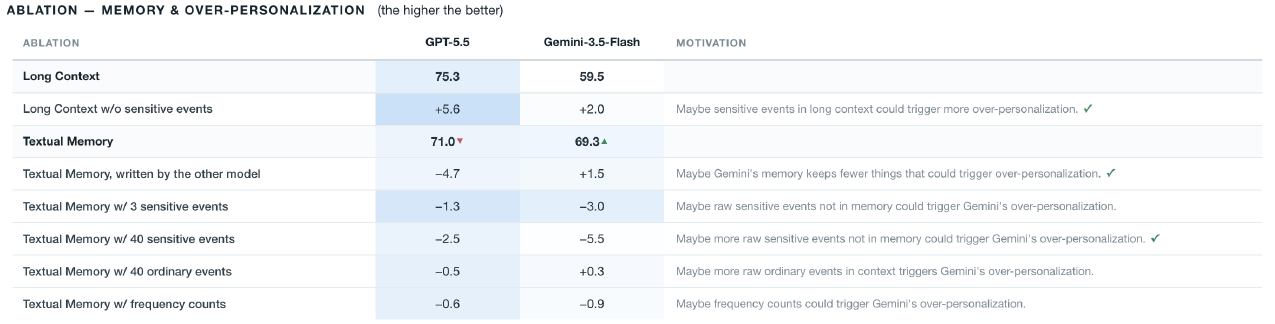}
    \caption{Memory and over-personalization ablations for GPT-5.5 and Gemini-3.5-Flash. Higher scores indicate stronger over-personalization against unnecessary, sensitive, repetitive, or socially awkward personalization. The table compares different assumptions why Gemini-3.5-Flash observes less over-personalization with a memory module, but opposite for GPT-5.5. We studies raw long context, long context without sensitive events, textual memory, memory written by the other model, textual memory augmented with sensitive or ordinary events, and memory with explicit frequency counts. We find that it is not simply whether a memory is good or bad, but that memory changes what becomes salient to the model’s attention and how each foundation model interprets it.}
    \label{fig:memory_ablation}
  \end{figure}%

To study this mechanism, we run the ablation experiment shown in Figure~\ref{fig:memory_ablation}. In our evaluation, we ask each model to write its own memory using the same memory-builder prompt. We deliberately use a simple, general prompt without over-engineering it: \textit{"You maintain a running memory of one user, updated as you read their activity in chronological chunks. Remember the useful parts of the user's persona and preferences from their activities across these apps; maintain the memory with ADD, EDIT, REMOVE, and MERGE actions; and the memory should contain only the resulting persona profile rather than a raw log of activity."} We find that it is not simply whether a memory is good or bad, but that \textbf{memory changes what becomes salient to the model's attention and how each foundation model interprets it}. For Gemini-3.5-Flash, raw events in the long-context prompt appear especially salient: the model repeatedly sees sensitive events and often pulls them into later answers too directly. Textual Memory does not necessarily remove those sensitive facts; it changes their form by compressing repeated private evidence into a bounded profile. Gemini-3.5-Flash appears better at treating those memory entries as background context or constraints rather than always-active personalization signals. In contrast, GPT-5.5 appears to treat textual memory as a clean user profile: once sensitive facts or strong preferences enter the memory, everything in that document becomes salient and safe-looking evidence to reuse. 

\begin{figure}[t]
    \centering
    \includegraphics[width=0.9\linewidth]{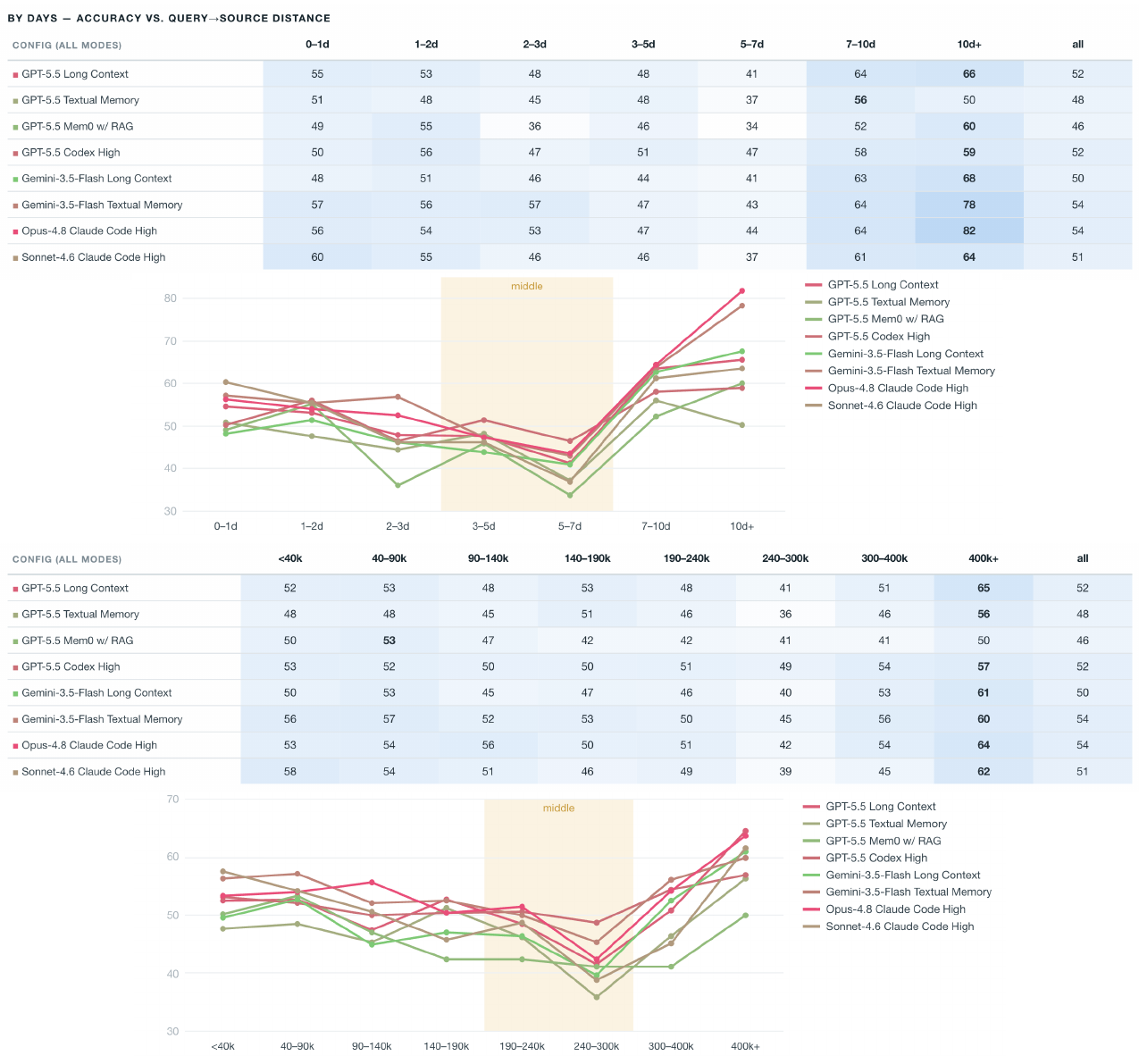}
    \caption{Lost-in-the-middle effects by temporal and token distance. We bin task accuracy by the distance from the earliest supporting evidence to the test query, measured both in days and number of tokens. Across model-mode configurations, performance often drops in the middle bins and rises again for very recent evidence or very old evidence that is likely to be repeatedly reinforced. This suggests that merely making evidence available through long context, memory, or agentic search does not guarantee that the model will use the right evidence at the right time.}
    \label{fig:middle}
  \end{figure}%

These observations lead to three questions, with results shown in Figure~\ref{fig:memory_ablation}. First, if raw sensitive events in the context make Gemini-3.5-Flash over-personalize, what happens when we add raw sensitive posts back as extra context alongside compact memory? Adding 40 sensitive posts reduces three-task restraint by 2.5 points for GPT-5.5 and 5.5 points for Gemini-3.5-Flash, while adding ordinary posts changes the score by less than one point. This supports the view that raw sensitive context is especially harmful for Gemini-3.5-Flash. Second, does Gemini-3.5-Flash improve because it avoids keeping sensitive events in its own memory? The memory-swap ablation answers no: GPT-5.5 drops from 74.8\% with its own memory to 66.4\% with Gemini-3.5-Flash's memory, while Gemini-3.5-Flash rises from 63.5\% with its own memory to 73.2\% with GPT-5.5's memory. Third, is Gemini-3.5-Flash less over-personalize with a memory because it does not count how frequently these sensitive events appeared in the history? We add frequency counts for these events into the memory, and instead, the count ablation changes the score by less than one point. We conclude that the model behavior on how they treat information in memory modules and long context differ, resulting in different over-personalization behaviors.

\begin{figure}[t]
    \centering
    \includegraphics[width=\linewidth]{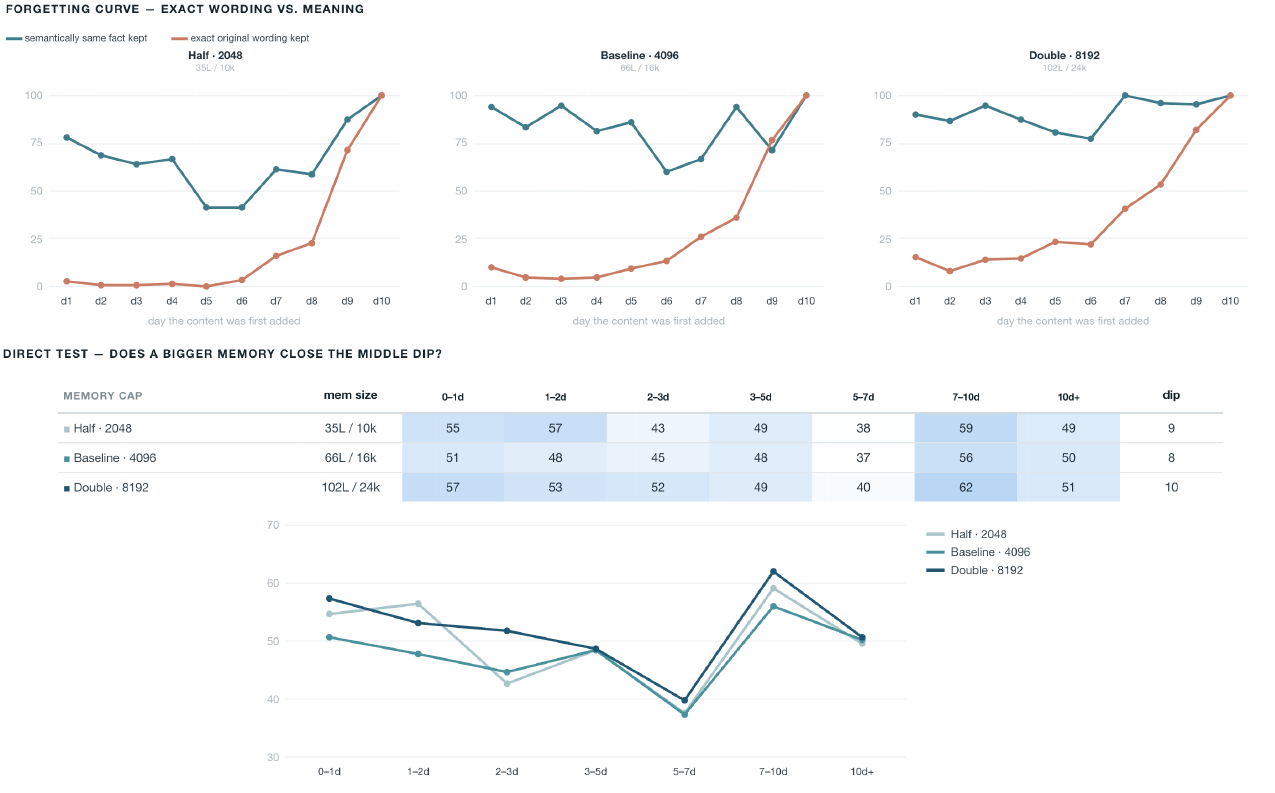}
    \caption{Memory forgetting curves under different Textual Memory capacity caps. The top row tracks whether content first added on each day remains in the final memory, either with the exact original wording or as a semantically equivalent fact, under 2048, 4096, and 8192 token caps. The bottom panel directly tests whether larger memory caps close the middle-position dip in downstream task accuracy. Larger caps preserve more semantically equivalent content, but exact wording remains fragile and the middle dip persists, suggesting that textual memory has its own self-evolution lost-in-the-middle failure mode.}
    \label{fig:forget_curve}
  \end{figure}%

\subsection{Lost-in-the-middle effects and memory forgetting curves}

We study how model performance changes with the distance from the earliest supporting evidence to the task query. As shown in Figure~\ref{fig:middle}, we bin this distance both by days and by intervening token volume. We acknowledge that due to real-world data's complexity, this is not a perfectly controlled position test: evidence that first appears early often has more chances to be reinforced by later activity, so the farthest bins may contain stable, repeated preferences rather than one isolated old fact. Even with this caveat, all model-mode configurations show a U-shaped pattern with accuracy drops in the middle over histories around 400k+ tokens. The middle region reflects a general long-context difficulty. Even when the relevant evidence exists somewhere in the available history, models are less reliable when that evidence sits neither very close to the query nor in the earliest, especially on personalization tasks beyond toy-examples like the needle-in-a-hay-stack task. 

We further study why memory does not simply remove the middle dip. Our assumption is that if the problem were only remembering a fact, a compact memory should preserve the useful concept even when the raw event sits in the middle of a long context. We first observe that automatic memory evolution rewrites memory aggressively: iterative memory drops or rephrases about 14.4\% of its lines at each update, and exact wording survival is even low, especially for content added early or in the middle of memory construction. However, this is not pure deletion. At the default 4096-token cap, concept-level survival is 83.1\%, meaning many facts remain semantically present even when their wording changes. This motivates the direct capacity question in Figure~\ref{fig:forget_curve}: if memory keeps more semantically equivalent content, does a larger memory close the middle dip? Increasing the GPT-5.5 Textual Memory cap from 2048 to 4096 to 8192 tokens increases the final memory from 35 to 66 to 102 bullet lines, and semantic concept survival rises from 63.2\% to 83.1\% to 90.8\% in average, with dip still in the middle of such memory's self-evolution. Overall accuracy changes modestly across all tasks, with the largest memory slightly above the baseline, with the middle dip remains 9\%, 7\%, and 10\% in accuracy, respectively. Therefore, \textbf{larger memory better preserves content over time, but textual memory still faces its own lost-in-the-middle problems during self-evolution process}.

\subsection{Data diversity}
\begin{figure}[t]
    \centering
    \includegraphics[width=\linewidth]{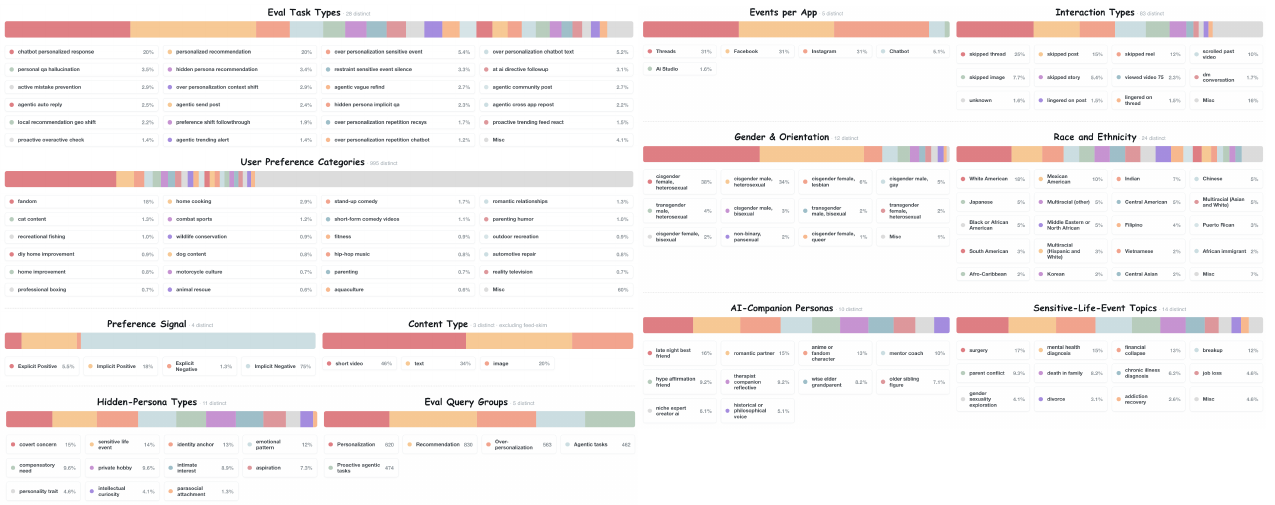}
    \caption{Data diversity in PersonaMem-v3. The benchmark spans heterogeneous task types, preference categories, preference signals, content formats, interaction patterns, hidden-persona motivations, sensitive life-event topics, demographic attributes, and AI-companion personas. This breadth lets the evaluation measure omni-platform user understanding across many kinds of personal evidence and social settings towards pluralistic alignment.}
    \label{fig:diversity}
  \end{figure}%

Figure~\ref{fig:diversity} summarizes the distributional coverage behind the released benchmark. The evaluation tasks cover personalization, recommendation, over-personalization, agentic tasks, and proactive agentic tasks, while the underlying user engagement histories include explicit and implicit preference signals, multiple content types, cross-platform interaction patterns, hidden personas, sensitive life events, demographic variation, and different AI-companion roles. This diversity is important for interpreting the results above: models must use personal context across heterogeneous evidence, privacy boundaries, and social situations towards pluralistic alignment.

\subsection{Error analysis}

\begin{figure}[t]
    \centering
    \includegraphics[width=0.9\linewidth]{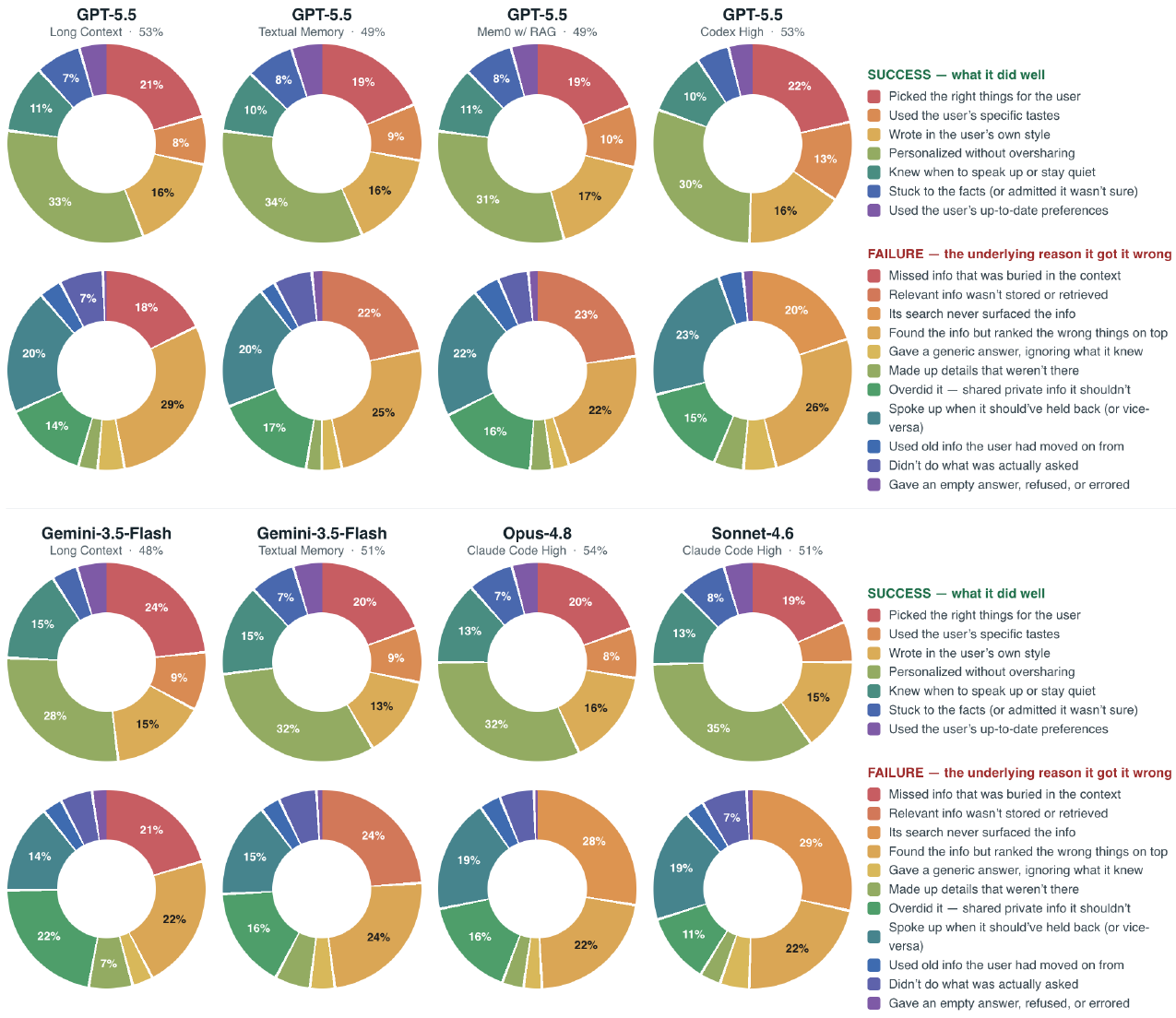}
    \caption{Qualitative success and failure taxonomy across model-mode configurations. Each pair of donut charts summarizes, for one configuration, what correct answers did well and what caused incorrect answers to fail. Successes cluster around selecting relevant items, using the user's concrete tastes, matching user voice, personalizing without oversharing, staying factual, and using up-to-date preferences. Failures separate buried-evidence misses, missing or unretrieved memory, candidate-ranking mistakes, generic answers, invented details, privacy leaks or over-withholding, stale preferences, off-task behavior, and empty or errored responses, showing that the main bottleneck is evidence judgment rather than context access alone.}
    \label{fig:error}
  \end{figure}%

The qualitative error analysis in Figure~\ref{fig:error} shows both what systems do well and why their wrong answers fail. The success rings are relatively similar: when a model is correct, it often succeeds by picking relevant items, matching the user's style, or using concrete tastes. The failure rings are more diagnostic. Retrieval errors mean the system fails to surface relevant past posts, messages, interactions, or preference evidence; ranking errors mean it finds plausible evidence or candidates but prioritizes the wrong one for the current personalization decision. Beyond these evidence-selection failures, systems also give generic answers, leak private or sensitive information, use stale preferences, make up details, act at the wrong time, or drift away from the requested task. Long Context has all evidence available, but still misses evidence buried in context in 18\% of its failures and often misprioritizes social media feed candidates. Textual Memory fails differently: 22\% of its failures are cases where the needed information was not stored or retrieved from the compressed memory. Mem0 with RAG has the same basic weakness at 23\%, and is especially vulnerable to stale preferences because a retrieved fact can be topically relevant while no longer being the right current guide. The Gemini-3.5-Flash rings also make the over-personalization pattern visible: Long Context has the largest oversharing slice at 22\%, while Textual Memory reduces that slice.

The Codex agentic harness has a distinctive profile. It is good at staying on task: 0\% of its failures are off-target cases where the agent answers a different request or centers the wrong topic. It also improves search-heavy tasks: GPT-5.5 rises from 16.9\% to 54.2\% on community voice drafting, from 27.5\% to 65.2\% on DM inbox digest, and from 48.8\% to 55.6\% on vague memory refinding. These tasks require retrieving multiple past conversational turns, direct messages, or user-authored posts, then summarizing the user's writing voice, inbox state, or remembered item rather than answering from a single obvious fact. Even after search, Codex failures are dominated by missed evidence or wrong prioritization among found candidates at 46\%, followed by proactive timing mistakes at 23\% and privacy over-application at 15\%, as shown in Figure~\ref{fig:error}. In the Codex agentic-task subset, 32.3\% of rows fall below the correctness cutoff; proactive daily catch-up and trend alert are especially weak, while direct send, repost, and reply actions are much stronger. The pattern suggests that Codex can keep the task in view, while still needing better evidence selection, privacy filtering, and final-answer calibration.

\begin{figure}[t]
    \centering
    \includegraphics[width=\linewidth]{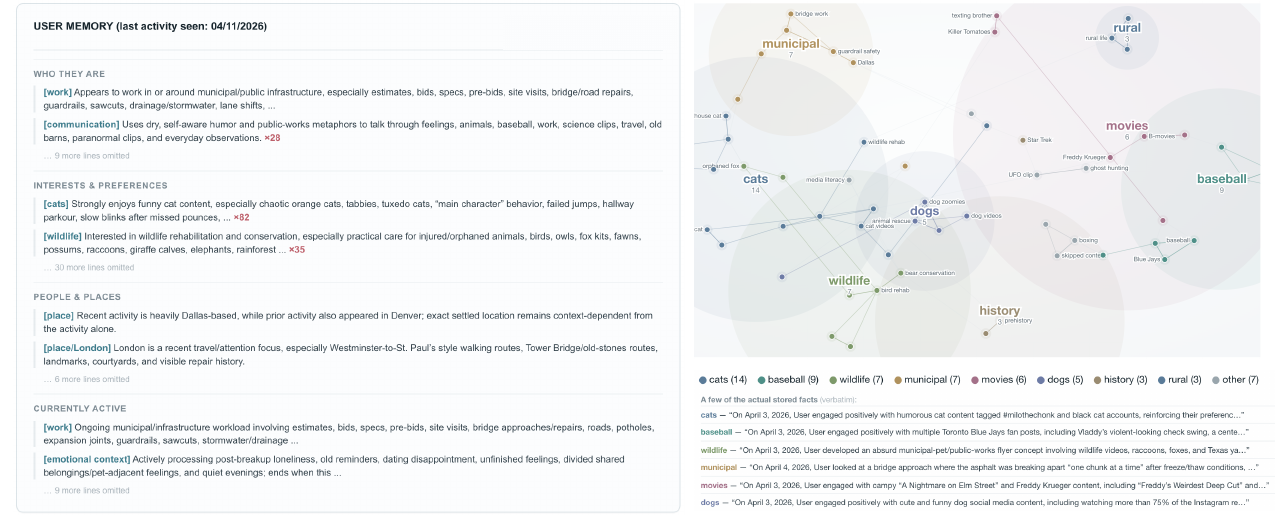}
    \caption{Examples of the two memory-based context-access modes. The left panel shows Textual Memory, a consolidated and inspectable user profile with sections such as who the user is, interests and preferences, people and places, and currently active context. The right panel shows a RAG-style vector memory implemented by Mem0 for one user, visualized with t-SNE clusters and example retrieved facts below its visualization. Textual Memory is compact and user-editable, while vector retrieval is targeted but can retrieve semantically close facts that miss timing, exact item details, user voice, or privacy constraints.}
    \label{fig:mem_rag}
  \end{figure}%

Claude Code fails differently from Codex. In the persona-1 trajectory audit, 28 agentic search trajectories average 2.36 turns, with search concentrated in vague refinding and group-DM summary rather than direct drafting rows. At the outcome level, Opus-4.8 and Sonnet-4.6 still drift off target in 6-7\% of failures and are dominated by missed evidence or wrong candidate ordering: 49\% for Opus-4.8 and 51\% for Sonnet-4.6. Opus-4.8 has a higher privacy-leak slice than Sonnet-4.6, while Sonnet-4.6 is especially prone to tidy answers whose ordering does not match the user's strongest evidence. Both Claude Code modes also under-act in proactive tasks, where there is often no concurrent user message. The prompt asks the agent to decide whether anything is worth proactively sending to the user, such as an unprompted short chat message, daily catch-up item, trending alert, or mistake-prevention warning, or whether it should stay silent. Claude Code often chooses silence when the benchmark expects a grounded proactive message, especially for close-friend updates, trend surfacing, and mistake-prevention alerts. Thus, agentic search improves access to evidence, while action timing and preference ordering remain separate modeling problems.

Together, these errors show that personal agents need more than larger context or more tool calls. The missing layer combines calibration, temporal awareness of user preferences, social intelligence, and ranking judgment: the agent must know whether evidence is current, whether acting would be welcome, and which candidate best fits the user's present context. These skills go beyond coding repositories, where tool use often has clearer ground-truth logic and less ambiguous social consequence. Besides, current coding harnesses with state-of-the-art models are not fully ready for perfect social-media feed recommendation, but Codex High reaches 35.3\% on proactive feed ranking, the highest among the evaluated settings as shown in Figure~\ref{fig:accuracy}, suggesting potential for agentic recommendation and reranking in the near future.

\subsection{Memory contents}

The two memory-based modes store different kinds of information, as illustrated by the memory and RAG examples in Figure~\ref{fig:mem_rag}. Textual Memory is a consolidated note written from the user's history. It contains sections such as who the user is, interests and preferences, people and places, and currently active context. This format is useful because it compresses thousands of events into a human-readable document that a model can scan cheaply. It is also more transparent to users than opaque latent representations or raw retrieval results, since users can inspect what the system has inferred about them, identify outdated or incorrect entries, and directly edit or correct the memory when needed.

Mem0 with RAG stores short dated facts rather than the user's posts directly, embeds those facts into a vector space, and retrieves a few nearest facts for each query. A single user's fact store contains clusters we visualize using t-SNE, such as pet videos, sports, wildlife rehabilitation, local infrastructure, unusual science clips, media-literacy work, and skipped content. This makes Mem0 cheap and targeted, and also explains its failure modes. A retrieved memory can be semantically close but temporally stale; it can describe a true preference but omit a later reversal; or it can retrieve a positive topic while missing the specific item, voice, or privacy constraint the task needs.


\subsection{Automated data-quality verification}
\label{subsec:auto_verify}

\begin{figure}[t]
    \centering
    \includegraphics[width=\linewidth]{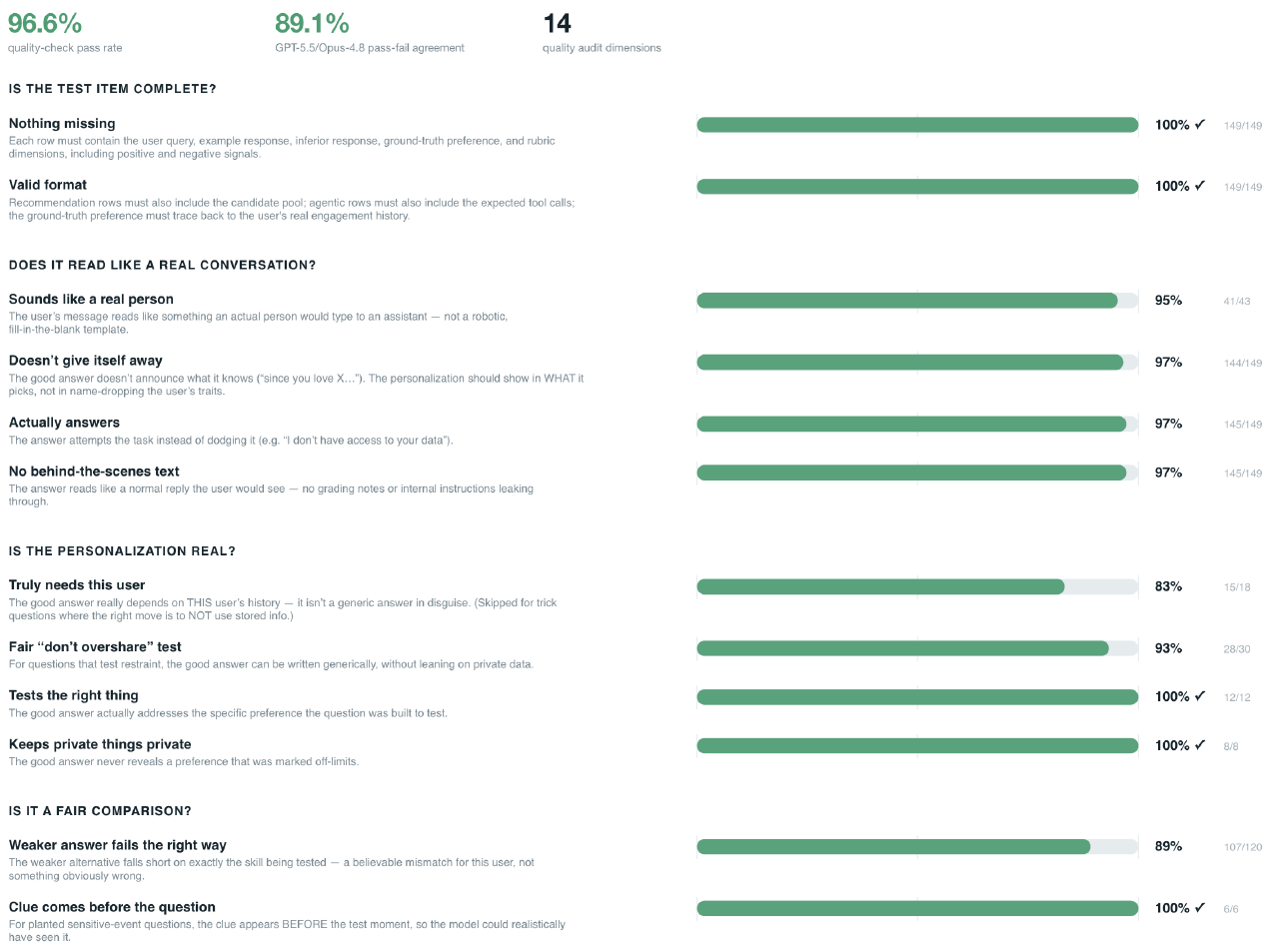}
    \caption{Post-hoc automated data-quality audit of released benchmark rows. GPT-5.5 re-checks each row under the same verification criteria used during generation, including completeness, valid format, conversational naturalness, dependence on user history, privacy restraint, and fairness of the inferior response. The released rows achieve a 96.6\% average pass rate across 14 quality dimensions, and GPT-5.5 and Opus-4.8 agree on pass-fail judgments 89.1\% of the time. This suggests that the benchmark quality filters are largely consistent across independent LLM judges rather than being an artifact of one verifier model.}
    \label{fig:quality}
  \end{figure}%

PersonaMem-v3 includes an automated quality-control pass for every benchmark query before release. The goal is to catch data artifacts that would make an evaluation instance misleading. This verification starts from the user model itself. As described in Section~\ref{subsubsec:atomic_inference} and \ref{subsec:psych_plumbing}, predicted preferences and personas are not kept just because one event suggests them. They must be supported by repeated evidence across the user's real engagement logs, confidence thresholds, and psychological anchors, so the benchmark does not build tasks from weakly grounded guesses. 

We then use GPT-5.5 as an LLM judge to check the generated benchmark examples. The verifier focuses on whether each row is a valid test of personalization, not an artifact. First, it checks the task setup: the query should not be answerable correctly without user context, and it should not leak the answer by explicitly pointing to the relevant preference. Second, it checks our golden example responses in the benchmark: the response should be natural, satisfy the universal personalization rubric in Section~\ref{subsec:rubric}, use the ground-truth preference or persona signal when needed, and use it implicitly and appropriately rather than with overly explicit framing such as “I know you like X.” Third, it checks the inferior response: the inferior should also be natural and comparable in length and format, but should differ from the example on the specific capability the task is meant to test, such as missing the relevant preference, using an outdated signal, leaking private context, ranking the wrong item, or over-personalizing, depending on the specific tasks. Both example and inferior responses shall be reasonable in general. Finally, the verifier checks the row format and evidence: each row must contain the user query, example response, inferior response, ground-truth preference, and rubric dimensions, including positive and negative signals; recommendation rows must also include the candidate pool; agentic rows must also include the expected tool calls. Besides, the ground-truth preference must trace back to the user's real engagement history.

A realistic personalization benchmark should not collapse into many versions of the same average user. In PersonaMem-v3, the primary grounding for each persona is still the user’s real-world activity history, with approximately 4,000 engagement events per user on average. These activities determine the preferences, hidden personas, and interaction patterns that the benchmark evaluates. However, some secondary attributes, such as names, broad education background, writing surface style, or companion-role framing we synthesize, may be under-specified by the raw data, and an LLM can repeatedly fill in these gaps with the same common defaults. To avoid this, PersonaMem-v3 adds a diversity control before generation: each user receives reproducible high-level anchors along several diversity dimensions, derived from the user identifier. These anchors help the cohort spread across broader demographic, psychological, social, and stylistic patterns. When the real activity history clearly contradicts an anchor, the data overrides it, if necessary, to ensure both diversity and real-world grounding.

If an example fails any of the quality filters above, the pipeline regenerates the flawed part, re-checks it, and removes the row if it still fails. Beyond these quantitative measurements, we also conduct qualitative human inspection ourselves to identify subtle data-quality issues and artifacts, iterating over a hundred of times. Each manual quality improvement is automatically summarized into a Markdown file, which preserves our subjective taste about data quality and provides reusable guidance for subsequent large-scale automated auditing post each generation. This makes data quality a built-in part of the benchmark construction process, ensuring that only samples passing all quality filters remain in the released benchmark. 

We perform further post-generation audit on data quality, as shown in Figure~\ref{fig:quality}. We use GPT-5.5 as an judge to re-evaluate the generated data quality under the same verification criteria, and observe an average quality pass rate of 96.6\%. To reduce the risk that this result overfits to a single judge model, we also run the same audit prompt with Opus-4.8. The two judges show 89.1\% agreement, suggesting that the high pass rate is not merely an artifact of one verifier model, but is largely consistent across independent LLM judges.

\section{Related Work}
\label{section:related_work}

PersonaMem-v3 is the third generation of the PersonaMem series, a progression toward building realistic and evaluable user histories for personal intelligence. Table~\ref{table:personamem_progression} summarizes how the series moves from fully synthetic chatbot histories in PersonaMem-v1 and PersonaMem-v2 to real-world-grounded, omni-platform user histories and agents in PersonaMem-v3.

{\footnotesize
\begin{longtable}{p{0.15\textwidth}p{0.25\textwidth}p{0.26\textwidth}p{0.26\textwidth}}
  \caption{Progression of user-history construction across the PersonaMem v1-v3 series.}
  \label{table:personamem_progression} \\
  \toprule
  \textbf{Dimension} & \textbf{PersonaMem-v1} & \textbf{PersonaMem-v2} & \textbf{PersonaMem-v3} \\
  \midrule
  \endfirsthead

  \toprule
  \textbf{Dimension} & \textbf{PersonaMem-v1} & \textbf{PersonaMem-v2} & \textbf{PersonaMem-v3} \\
  \midrule
  \endhead

  \midrule
  \endfoot

  \bottomrule
  \endlastfoot

  Data source & 20 fully synthetic users & 1000 fully synthetic users with more comprehensive personas & 200 anonymized \textbf{real-world} users with 4,000,000 engagement histories \\[2pt]
  Explicit vs.\ implicit & Explicit user preferences & \textbf{Implicit} user preferences & Around 95\% \textbf{implicit} user behavior signals \\[2pt]
  Scenarios & Chatbot conversations & Chatbot conversations & \textbf{Omni-platform}, including chatbot, social media recommendation, \textbf{agentic tasks}, and proactiveness \\[2pt]
  Restraint & Personalization & Personalization & Personalization and \textbf{over-personalization} \\[2pt]
  User privacy & No mentioning of user private information & Including personally identifiable information and user-initiated ask-to-forget scenarios & Including psychology-anchored hidden persona and \textbf{socially inappropriate} scenarios \\[2pt]
  Dynamics & Fully synthesized preference updates & Fully synthesized preference updates & Reinforced, emerging, diminishing, bursting, and varied attention shifts from the \textbf{real world} \\
\end{longtable}
}

Beyond this within-series progression, we situate PersonaMem-v3 at the intersection of four broader lines of work: LLM and agentic memory, LLM personalization, LLM-powered recommendation, and personal agents. The literature in each area is growing quickly, but most benchmarks still isolate one capability at a time. Building on the omni-platform grounding introduced above, PersonaMem-v3 instead evaluates omni-platform personal intelligence: how a system builds user understanding from fragmented cross-platform evidence, uses that understanding for recommendation and agentic assistance, and balance personalization vs over-personalization.

\subsection{LLM and agentic memory}

Long-term memory has become a central systems layer for LLM assistants and agents. Benchmarks such as LongMemEval and LoCoMo evaluate whether systems can recover and reason over information from sustained interaction histories, including factual recall, temporal reasoning, knowledge updates, and long-range conversational consistency \citep{wu2024longmemeval,maharana2024locomo}. More recent memory-oriented evaluations move closer to agent settings: MemoryAgentBench studies incremental multi-turn memory under retrieval, test-time learning, long-range understanding, and selective forgetting, while EvoArena evaluates how agents and memories track evolving environments rather than static histories \citep{hu2025memoryagentbench,xu2026evoarena}.

In parallel, recent work has built explicit memory systems for LLM agents. MemoryBank introduces long-term memory for sustained companion-style interaction~\citep{zhong2024memorybank}. Mem0 studies production-oriented extraction, consolidation, and retrieval for scalable long-term memory~\citep{chhikara2025mem0}. A-MEM explores agentic memory organization through dynamically linked notes rather than fixed memory schemas~\citep{xu2025amem}. MIRIX extends memory beyond plain text to a multi-agent, multimodal memory system~\citep{wang2025mirix}. MemOS pushes the systems perspective further by treating memory as a first-class operating-system resource for AI systems~\citep{li2025memos}. A newer line studies self-evolving memory, where the memory system learns from deployment-time experience rather than only storing and retrieving facts. Memory-R1 and MemAgent train memory-management and answer-selection agents with reinforcement learning, learning memory operations automatically~\citep{yan2025memoryr1, yu2025memagent}. ReasoningBank distills reusable reasoning strategies from self-judged successful and failed trajectories, and uses memory-aware test-time scaling to improve future agent behavior~\citep{ouyang2026reasoningbank}. Evo-Memory benchmarks self-evolving memory over sequential task streams and proposes ReMem, which integrates reasoning, action, and memory updates after each interaction~\citep{wei2025evomemory}. EvoMemBench further standardizes memory evaluation from a self-evolving perspective by separating in-episode versus cross-episode memory and knowledge-oriented versus execution-oriented memory~\citep{wang2026evomembench}.

These works establish memory as a first-class capability, but their main question is usually whether information can be stored, retrieved, updated, or forgotten. PersonaMem-v3 evaluates memory as only one component inside omni-platform personal intelligence. The evidence for a user is fragmented across social media, chatbot sessions, calendar events, and geo-temporal context, with much of it is implicit, negative, evolving, or privacy-sensitive, and PersonaMem-v3 extends memorization towards personalization tasks.

\subsection{Personalization and over-personalization}

A large body of work studies personalization through user profiles, preference following, and personalized generation. Early work like LaMP introduces profile-conditioned personalized classification and generation, while LongLaMP and LaMP-QA extended this line to longer-form generation and question answering \citep{salemi2024lamp,kumar2024longlamp,salemi2025lampqa}. PersonalLLM studies heterogeneous user preferences through simulated users and reward-model-based preference variation \citep{zollo2024personalllm}. PrefEval, PersonaLens, and PersonaFeedback evaluate whether systems can infer, remember, and apply user-specific preferences in conversational or task-oriented settings \citep{zhao2025prefeval,zhao2025personalens,tao2025personafeedback}.

A separate line asks not only whether a model can personalize, but whether it personalizes appropriately. OP-Bench formalizes over-personalization in memory-augmented assistants as irrelevance, repetition, and sycophancy \citep{hu2026op}. RPEval studies rational preference utilization, showing that irrelevant personalized memories can interfere with intent understanding \citep{feng2026rpeval}.

PersonaMem-v1 and PersonaMem-v2 move closer to the setting of this paper. PersonaMem-v1 studies dynamic user profiling and personalized responses over evolving multi-session histories \citep{jiang2025know}. PersonaMem-v2 focuses on implicit preferences revealed through ordinary user-chatbot interactions and studies reinforcement fine-tuning and agentic memory as scalable personalization mechanisms \citep{jiang2025personamem}. These benchmarks are important precedents for temporal and implicit personalization, but they still operate primarily within a conversational channel.

Compared with prior personalization benchmarks, PersonaMem-v3 expands both grounding and scope. It is grounded in million-scale anonymized engagement histories rather than only simulated profiles or single-platform dialogues, and it evaluates personalization jointly with recommendation, over-personalization, personalized agentic tasks, proactiveness, temporal preference tracking, and privacy-sensitive restraint.

\subsection{LLM-powered recommendation}

A growing line of work treats LLMs as rerankers or reasoning layers on top of existing recommender pipelines. Zero-shot ranking work showed that LLMs can rerank candidate sets without retraining the full recommender stack \citep{hou2024zeroshotrankers}. LLM4Rerank further studies LLM-based reranking under multiple objectives such as accuracy, diversity, and fairness \citep{gao2025llm4rerank}. RecRanker uses instruction tuning to make LLMs act as top-$k$ rankers \citep{luo2025recranker}. Parallel industrial work studies LLMs as tools for user-interest exploration and feedback alignment, using language models to broaden or steer recommendations beyond immediate historical loops \citep{wang2024interestexploration,wang2025userfeedbackalignment}.

Another line moves toward end-to-end generative or agentic recommenders. OneRec proposes an end-to-end generative recommendation paradigm at industrial scale \citep{zhou2025onerec}. OpenOneRec releases an open technical stack, benchmark, and foundation models for recommendation-oriented generative modeling \citep{zhou2025openonerec}. OneReason studies how to activate explicit reasoning in generative recommendation \citep{onereason2026technical}. AgenticRec optimizes the full tool-use and ranking trajectory for recommender agents \citep{li2026agenticrec}. MARS adds a hierarchical belief-state memory that explicitly models preference lifecycle dynamics for agentic recommendation \citep{shen2026mars}.

PersonaMem-v3 complements rather than replaces this literature. It does not assume an LLM should own the full recommender pipeline. Instead, it evaluates whether an LLM can serve as a personalized reranker and assistant on top of candidate pools, using cross-platform user understanding and near-future engagement ground truth. Its recommendation tasks are also embedded in a broader personal-agent setting that jointly measures recommendation quality, user steering, hidden-persona alignment, and over-personalization in recommendation, not as an isolated ranking problem.

\subsection{Personal intelligent agents}

Recent work increasingly evaluates agents as persistent collaborators rather than static task solvers. VitaBench 2.0 studies personalized and proactive agents over temporally ordered long-term user interactions, where preferences are embedded in fragmented heterogeneous records \citep{chen2026vitabench2}. ProAgentBench decomposes proactive assistance into timing prediction and assist-content generation, using real working-session data rather than purely synthetic trajectories \citep{tang2026proagentbench}. ContextAgent studies proactive assistants that act on open-world sensory context \citep{yang2025contextagent}. PersonalAlign focuses on personalized GUI agents that resolve implicit intent by reasoning over long-term user records \citep{lyu2026personalalign}.

Industry systems and technical reports show the same shift from chat to persistent, tool-connected personal agents. Google's Project Astra frames the long-term product vision as a universal AI assistant that can understand live context, use memory, share screens, process video, and eventually take action across devices~\citep{google2025projectastra}. Gemini Spark makes this more explicit as a 24/7 personal AI agent that works in the background across Gmail, Calendar, Drive, Docs, Sheets, Slides, YouTube, Maps, and web tasks, with user direction and confirmation before major actions~\citep{google2026geminispark}. Meta's Muse Spark is a natively multimodal reasoning model with tool use, visual chain of thought, and multi-agent orchestration, powering Meta AI across apps and glasses on the path toward personal superintelligence~\citep{meta2026musespark}. OpenAI's Operator and ChatGPT agent similarly combine browser interaction, research workflows, external data access, code execution, connectors, and action execution in user-facing agent stacks~\citep{openai2025agentsplatform, openai2025chatgptagent, openai2025operator}. Anthropic's Claude Code and the Model Context Protocol (MCP) emphasize persistent tool access, file and shell operations, and standardized connections between agents and external systems~\citep{anthropic2024mcp, anthropic2026claudecode}. OpenClaw represents a different deployment point: a self-hosted, open-source personal assistant with heartbeat that connects to messaging apps, browsers, files, shell commands, skills, plugins, and persistent memory~\citep{openclaw2026}. Hermes Agent similarly targets persistent personal agency, with cross-channel interfaces, long-term memory, session search, autonomous skill creation, subagents, and self-improving skills~\citep{nousresearch2026hermesagent}. OpenJarvis studies personal AI on personal devices and explicitly frames systems such as OpenClaw and Hermes Agent as emerging personal AI stacks, decomposing them into intelligence, engine, agents, tools and memory, and learning primitives~\citep{saadfalcon2026openjarvis}.

PersonaMem-v3 is close in spirit to this agent literature, but contributes an open-source data and model specification derived from million-scale anonymized real-world user activity, targeting omni-platform personal intelligent agents together with a comprehensive evaluation harness.

\section{Conclusion}

PersonaMem-v3 shows that omni-platform personal intelligence is an open and under-measured capability frontier. By grounding user worlds in million-scale anonymized engagement histories and extending them across social media, chatbot, companion-chat, calendar, and geo-locations over time, the benchmark evaluates personalization as a holistic user understanding problem rather than a single-app recommendation or memory-retrieval task, together with fine-grained preference evolution from real-world records. Our results show that even the strongest evaluated agents solve only slightly more than half of the benchmark, revealing substantial gaps in cross-platform personalization, over-personalization, neural recommendation, proactiveness, temporal preference tracking, personalized agentic tasks, and socially appropriate restraint.

Across different evaluation modes, we find that long context, textual memory, vector-based memory retrieval, and agentic search each offer different tradeoffs. Long context exposes the raw evidence but remains costly and vulnerable to lost-in-the-middle failures. Textual Memory and Mem0 greatly improve efficiency and preserve broad, stable preferences, but they can also omit exact details, stale or negative evidence, and privacy constraints. Agentic search gives models a more realistic way to inspect structured user databases through agentic search and tool calls, and is especially useful for evidence-grounded tasks such as refinding past content, summarizing message threads, drafting in the user’s community voice, and acting across platforms. However, tool access alone does not solve the harder parts of personal intelligence: the agent still has to decide what evidence matters, whether it is still valid, how much to personalize, whether an action would be socially appropriate, and when it should stay silent. In particular, agentic harnesses improve evidence access for search-heavy tasks, but remain weak on recommendation reranking, proactive timing, mistake-prevention alerts, and hidden-persona reasoning, where the bottleneck is judgment rather than retrieval.

Moreover, the lost-in-the-middle problem is not limited to long-context reasoning, but can also appear in the self-evolving process of textual memory: as memory is incrementally updated over time, information from middle segments can become more vulnerable to being overwritten, compressed away, or under-emphasized in later memory states. Adding a memory module also changes what becomes salient to the model and how each foundation model interprets that salience. For example, GPT-5.5 pays a larger over-personalization tax when equipped with Textual Memory, suggesting that it may treat the memory as a clean user profile whose contents are safe to reuse. Gemini-3.5-Flash shows the opposite pattern: although it has more severe over-personalization problems overall than GPT-5.5, Opus-4.8, and Sonnet-4.6, its Textual Memory setting mitigates over-personalization relative to raw long context, especially on sensitive-user-information cases. This suggests that memory is not merely a compression mechanism. It also reshapes the model’s attention, evidence prioritization, and privacy behavior in model-specific ways.

Looking forward, these findings suggest that the next generation of personal intelligent agents should go beyond storing memories, retrieving evidence, or calling tools. They must decide which user signals are relevant to the current moment, which have expired or been contradicted, which should remain private, and whether a response or action would actually be helpful. This requires memory and agentic policies that model recency, preference evolution, negative evidence, privacy boundaries, social appropriateness, and action timing directly. We hope PersonaMem-v3 provides a realistic evaluation foundation for building personal agents that understand users across digital platforms for holistic user understanding, recommendation, and agentic tasks, while using personal context responsibly and comfortably.

\section{Limitations}
\label{section}

PersonaMem-v3 is designed for privacy-preserving release, which necessarily shapes the form of the benchmark. Since original multimodal user data and raw social-media posts cannot be released, images, videos, and platform activities are represented through textual annotations and structured metadata rather than native media or internal platform records. Generating multimodal data at this scale is also too unrealistic and expensive. This design enables an open, inspectable, and reproducible benchmark while still preserving many key personalization signals across modalities and platforms.

The psychological and social-linguistic frameworks used in PersonaMem-v3 are intended as generation and evaluation scaffolds rather than diagnostic claims. Traits, motives, coping patterns, audience design, self-presentation, and parasocial attachment provide useful structure for making hidden preferences, social context, and restraint requirements more explicit and auditable. However, the benchmark does not aim to validate clinical labels or prove any psychological theory as a complete account of user behavior.

Finally, PersonaMem-v3 focuses on Meta-family platforms and chatbot interactions, and does not cover every possible digital footprint, private platform MCP interface, or production database schema. Nevertheless, this setting already captures a substantially richer omni-platform personalization environment than prior open-source benchmarks, spanning memory, recommendation, social context, and agentic action. We hope this release helps motivate future work on longer-running personal agents with agentic coding and tool call capabilities and richer multimodal and multilingual environments.

\clearpage
\newpage
\bibliographystyle{assets/plainnat}
\bibliography{paper}


\end{document}